\documentclass[sigconf,nonacm]{acmart}
\usepackage{lipsum}
\usepackage{tikz}
\usepackage{amsmath}
\usepackage{graphicx}
\usepackage{xspace,soul}
\usepackage{algorithmic}
\usepackage{tabularx}
\usepackage{harveyballs}
\usepackage{fancyhdr}
\usepackage{comment}
\usepackage{dblfloatfix}    % To enable figures at the bottom of page
\usepackage{multirow}
\usepackage{subfigure}
\usepackage[utf8]{inputenc}
\usepackage[flushleft]{threeparttable}
\usepackage{longtable}
\usepackage{pifont}
\usepackage{xcolor}
\usepackage{tikz}
\usepackage{graphicx}
\usepackage{listings}
\usepackage{circledsteps}
\usepackage{array,booktabs}

\pgfkeys{/csteps/inner color=white}
\pgfkeys{/csteps/fill color=black}

\definecolor{codegreen}{rgb}{0,0.6,0}
\definecolor{codegray}{rgb}{0.5,0.5,0.5}
\definecolor{codepurple}{rgb}{0.58,0,0.82}
\definecolor{backcolour}{rgb}{0.95,0.95,0.92}

\definecolor{lightgray}{gray}{0.8}

\lstdefinestyle{mystyle}{
    backgroundcolor=\color{backcolour},   
    commentstyle=\color{codegreen},
    keywordstyle=\color{magenta},
    numberstyle=\tiny\color{codegray},
    stringstyle=\color{codepurple},
    basicstyle=\ttfamily\footnotesize,
    breakatwhitespace=false,         
    breaklines=true,                 
    captionpos=b,                    
    keepspaces=true,                 
    numbers=left,                    
    numbersep=5pt,                  
    showspaces=false,                
    showstringspaces=false,
    showtabs=false,                  
    tabsize=2
}
 
\def\BibTeX{{\rm B\kern-.05em{\sc i\kern-.025em b}\kern-.08em
    T\kern-.1667em\lower.7ex\hbox{E}\kern-.125emX}}

\newcommand{\system}{FP-Rowhammer\xspace}

\begin{document}

\title{\system: DRAM-Based Device Fingerprinting}
%\author{Anonymous authors}
\author{Hari Venugopalan}
\email{hvenugopalan@ucdavis.edu}
\affiliation{
\institution{UC Davis}
\country{ }
}

\author{Kaustav Goswami}
\email{kggoswami@ucdavis.edu}
\affiliation{
\institution{UC Davis}
\country{ }
}

\author{Zainul Abi Din}
\email{zdin@ucdavis.edu}
\affiliation{
\institution{UC Davis}
\country{ }
}

\author{Jason Lowe-Power}
\email{jlowepower@ucdavis.edu}
\affiliation{
\institution{UC Davis}
\country{ }
}

\author{Samuel T. King}
\email{kingst@ucdavis.edu}
\affiliation{
\institution{UC Davis}
\country{ }
}

\author{Zubair Shafiq}
\email{zubair@ucdavis.edu}
\affiliation{
\institution{UC Davis}
\country{ }
}
\begin{abstract}
Device fingerprinting leverages attributes that capture heterogeneity in hardware and software configurations to extract unique and stable fingerprints. 
Fingerprinting countermeasures attempt to either present a uniform fingerprint across different devices through normalization or present different fingerprints for the same device each time through obfuscation.
We present \system, a Rowhammer-based device fingerprinting approach that can build unique and stable fingerprints even across devices with normalized or obfuscated hardware and software configurations. 
To this end, \system leverages the DRAM manufacturing process variation that gives rise to unique distributions of Rowhammer-induced bit flips across different DRAM modules.
%
%\system's design and implementation is able to overcome memory allocation constrains without requiring root privileges.
%
Our evaluation on a test bed of 98 DRAM modules shows that \system achieves 99.91\% fingerprinting accuracy. 
\system's fingerprints are also stable, with no degradation in fingerprinting accuracy over a period of ten days.
We also demonstrate that \system is efficient, taking less than five seconds to extract a fingerprint.
\system is the first Rowhammer fingerprinting approach that is able to extract unique and stable fingerprints efficiently and at scale.
\end{abstract}

\maketitle

\section{Introduction}
\label{sec:intro}
Growing restrictions against stateful identifiers \cite{idfa, adid, cookie} has led to
the emergence of device fingerprinting for
tracking \cite{laperdrix2020browser,iqbal2021fingerprinting, iosfingerprinting}.
Fingerprinters capture distinguishing hardware and software attributes of devices to construct a \textit{fingerprint} that identifies clients without
needing to store any client-side state \cite{laperdrix2020browser}.
%Stateless tracking is becoming more prevalent \cite{iqbal2021fingerprinting, iosfingerprinting} in response to emerging countermeasures against stateful tracking  \cite{idfa, adid, cookie}.
%
%Stateless tracking relies on capturing distinguishing  hardware and software attributes of a device to construct a \textit{fingerprint} without 
%needing to store any client-side state \cite{laperdrix2020browser}.
%
For a fingerprint to be useful, it needs to be \emph{unique} and \emph{stable}.
First, a fingerprint should have sufficiently high entropy to uniquely identify a device within a population of devices 
\cite{browser-uniqueness}. 
Second, it should remain sufficiently
stable over time so it can be linked to previous fingerprints from the same device for re-identification \cite{fp-stalker}.

Fingerprinting techniques typically aggregate a variety of software and hardware 
attributes, such as screen resolution, the number of processors, and the type and version
of the operating system, to construct device fingerprints \cite{fingerprintjs-github, fpjs-android, fpjs-ios}. 
To attain high entropy, such fingerprints are dependent on devices having diverse configurations that are sufficiently distinguishable. 
A common countermeasure to reduce entropy is to  normalize the attributes that capture device 
configurations to present the same values across different devices \cite{tor, brave-canvas}.
Fingerprinters also have to contend with how the attributes change over time with the goal of producing a stable fingerprint.
These changes can either arise from natural evolution of device configurations (e.g., software updates) or 
from fingerprinting countermeasures that intentionally obfuscate attributes \cite{farbling, canvas-defender}. 
To attain high stability, fingerprinters attempt to predict fingerprint changes \cite{fp-stalker} or employ stemming to improve stability \cite{stemming}.

In this work, we investigate the threat model where a fingerprinter aims to extract unique and stable fingerprints for devices with identical hardware and software configurations over an extended time period.
To this end, we aim to capture fundamental differences in the physical properties of the device's hardware. 
The key insight is that a fingerprinter can capture inherent differences that arise as a result of \textit{process variation} in the hardware manufacturing process. 
As users rarely modify their device hardware, these fingerprints would remain stable. 
While prior research has exploited the manufacturing process variation in CPUs \cite{demicpu}, GPUs \cite{drawn_apart-gpu}, and clocks \cite{clock-skew}, we successfully employ memory (specifically DRAM) for device fingerprinting.

We leverage Rowhammer \cite{kim-rowhammer} to extract fingerprints by capturing the side-effects of process variation in memory modules. 
At a high level, ``hammering'' a memory row (i.e., repeated read or write operations in a short time interval) results in bit flips in adjacent memory rows. 
In this paper, we investigate whether the pattern of bit flips due to Rowhammer can be leveraged for fingerprinting.
To build intuition, Figure~\ref{fig:starry_sky} visualizes the distribution of bit flips produced by executing Rowhammer at the same locations on two identical DRAM modules (also called Dual Inline Memory Modules or DIMMs) at two different points in time.
The figure shows that the distribution of bit flips is \textit{reasonably similar} on the same DRAM modules at different points in time while being \textit{noticeably different} across different DRAM modules.
While recent research has extended Rowhammer-based PUFs \cite{rowhammer-puf} for 
fingerprinting \cite{fp-hammer}, as we discuss next, there are several non-trivial, 
technical concerns that make it challenging to use Rowhammer for fingerprinting 
\cite{rowhammer-reproduction}.

We present \system, a Rowhammer-based fingerprinting approach that overcomes these challenges and exploits bit flip distributions to extract \textit{unique} and \textit{stable} fingerprints even among homogeneous devices with identical software and hardware configurations over an extended period of time. 
We address the following technical challenges with \system:
\begin{itemize}
\item First, we find that the bit flips triggered by Rowhammer are 
non-deterministic (i.e., hammering the same memory location does not 
always flip the same set of bits). 
In \system, we use statistical sampling to account for \textbf{non-determinism} by hammering the same memory locations multiple times to extract and compare probability distributions of bit flips. This leads to accurate and robust fingerprinting, 
even in presence of external factors ($\S$ \ref{app:eval_baseline}).

\begin{figure}[t]
\centering
  \includegraphics[width=.8\columnwidth]{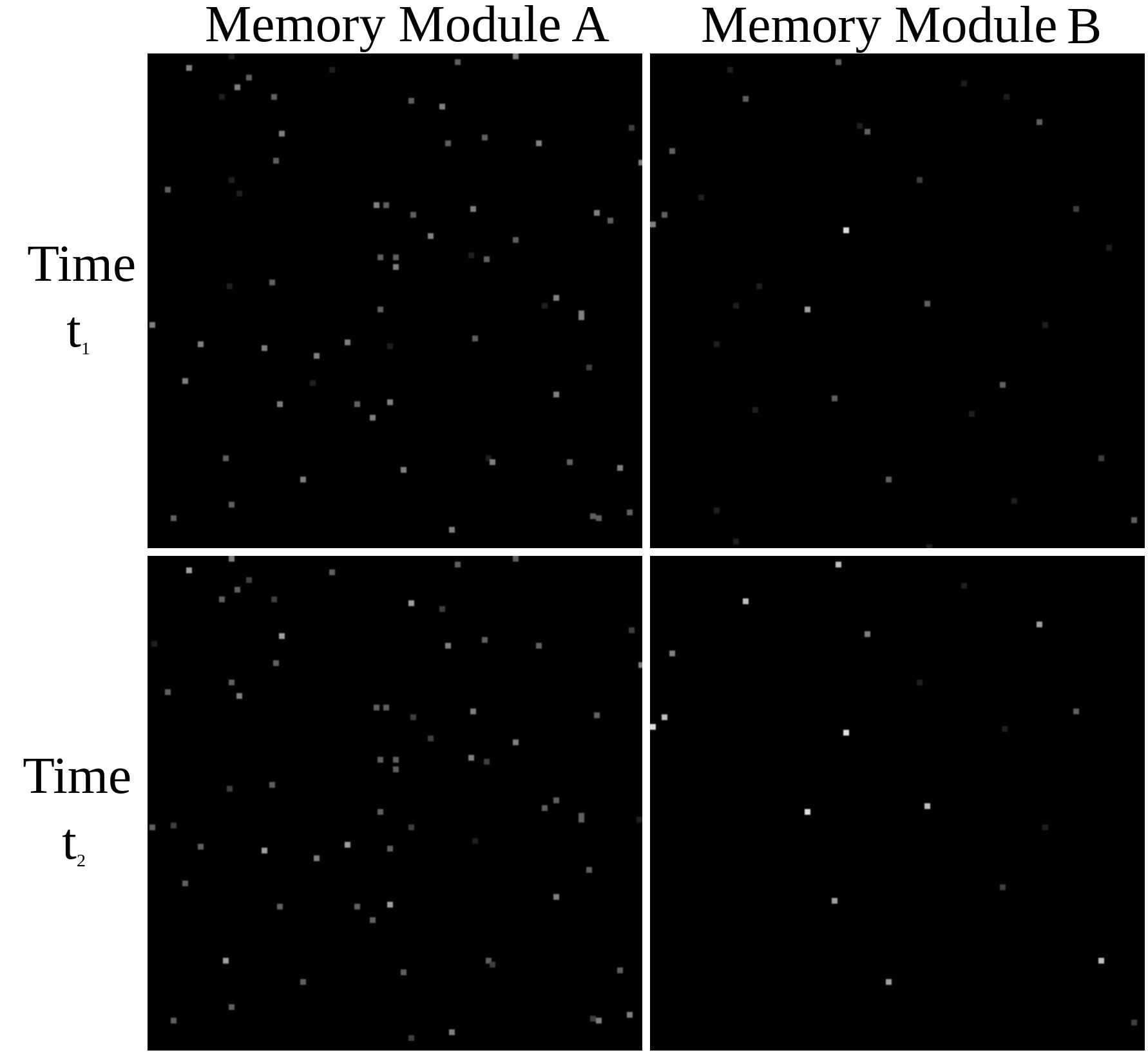}
  \vspace{-3mm}
  \caption{Visualization of the Rowhammer bit flip distribution with a 
  brighter spot representing a higher bit flip probability. 
  The \textbf{top row} shows the distributions on two different but 
  identical DRAM modules. The \textbf{bottom row} shows the distribution 
  on the same DRAM modules at a later point in time. It shows that the distribution of bit flips is \textit{reasonably similar} on the same DRAM modules at different points in time while being \textit{noticeably different} across different DRAM modules.}
  \label{fig:starry_sky}
\end{figure}
\item Second, the operating system's \textbf{memory 
abstractions limit access} to contiguous
physical memory. Since bit flips are not uniformly distributed across rows \cite{rowhammer-reproduction},
fingerprinters will struggle to match fingerprints if these abstractions present them with different chunks
of memory at different points in time.
In \system, we sample chunks of memory for near-guaranteed access to the same chunk.
\item Third, memory modules implement \textbf{Rowhammer mitigations}, such as Target Row Refresh (TRR) \cite{micron-trr}. 
While many-sided and non-uniform hammering can bypass mitigations \cite{trrespass, blacksmith}, we find that
certain patterns are not suitable for fingerprinting ($\S$\ref{sec:eval_freq}). In \system, we 
systematically identify hammering patterns to bypass these mitigations for at-scale 
Rowhammer-based device fingerprinting.
\end{itemize}

\vspace{-1mm}
We evaluate \system on a test bed of nearly 100 DIMMs across 6 sets of identical DRAM modules for 2 major DRAM manufacturers with the largest DRAM marketshare.
\system achieves a high fingerprint accuracy of 99.91\%, corresponding to 100\% precision and 97.06\% recall. 
\system also achieves high stability, with no degradation in fingerprint accuracy over a period of ten days.
\system is able to efficiently extract fingerprints in less than five seconds.
To the best of our knowledge, our evaluation of DRAM modules is the largest scale to date.

\begin{comment}

Our key contributions include:

\begin{itemize}

\item Handling non-deterministic bit flips: We handle non-deterministic bit flips by hammering the same memory chunks multiple times and using the divergence between probability distributions of bit flips to re-identify devices.

\item Overcoming memory allocation constraints: We overcome memory allocation constraints by devising a novel sampling strategy that guarantees access to the same chunk of memory for fingerprinting.

\item Operationalizing bypass techniques for Rowhammer mitigations: We bypass Rowhammer mitigations by identifying effective hammering patterns that can trigger bit flips at-scale.

\end{itemize}

\end{comment}
\section{Background}
    \label{sec:background}    
    
    \subsection{DRAM basics} 
        \label{subsec:organization}
        \label{subsec:working-dram}        
        \label{sec:dram_commands}
    \iffalse
    \begin{figure}[!t]
    \centering
      \includegraphics[width=\columnwidth]{figs/dram_arch.jpeg}
      \caption{This figure shows a single rank of a DRAM DIMM. Each rank contains 
                multiple logical structures called banks that are interspersed 
                across multiple physical structures called chips. Each bank is 
                an array of cells in the form of rows and columns.\\}
      \label{fig:dram_arch}
    \end{figure}
    \fi
        All DRAM technologies follow the same basic architecture
        ~\cite{ddr2,ddr3,ddr4,ddr5,gddr5,gddr6,microLPDDRx,ddr3l}.
        We focus on DIMM-based DRAM packages in this paper, but our findings also apply to other packaging techniques.
        
        Each physical DIMM is installed on a DRAM channel on the motherboard. 
        Channels enable issuing concurrent requests to multiple DIMMs. 
        A DIMM can have one or more ranks. 
        Each rank contains multiple logical structures 
         called banks. A bank is a two-dimensional array of 
        cells organized into rows and columns. Each cell contains a capacitor 
        and an access transistor, with the capacitor's charged state representing 
        a single bit. The number of cells in a column is given by
        the width (x8, x16 etc) of the DIMM. %Figure \ref{fig:dram_arch}
        %visualizes one rank of a DIMM.
        
        The memory controller issues commands to the DRAM to perform memory operations at the granualrity of a \emph{row}. The \texttt{ACT} command 
        activates a row by loading it into the row-buffer before reading or writing to it. The \texttt{PRE} command deactivates a row and prepares the row-buffer to load another row by restoring previously held values. The memory controller
        periodically issues \texttt{REFI} commands to refresh
        the charge held by capacitors since the charge naturally drains
        over time. %Every DRAM capacitor typically gets refreshed at least
        %once every 64 milliseconds.
   
    \subsection{Rowhammer}
        \label{sec:rh}
        DIMMs are susceptible to memory corruption as a result of 
        electrical interference. Rowhammer \cite{intel-patent, kim-rowhammer} 
        corrupts the data stored in some capacitors leading to bit flips in memory. Specifically, Rowhammer triggers bit flips at a particular address by repeatedly accessing neighboring addresses. This leads to the repeated activation and deactivation of rows containing the accessed addresses. The resulting electro-magnetic 
        interference between the accessed rows (referred to as aggressors or aggressor 
        rows) and their neighboring rows (referred to as victim rows) accelerates the 
        charge dissipation of the capacitors in the victim rows, resulting in memory corruption. %Once 
        %these capacitors have lost a sufficient amount of charge, refreshing the DRAM 
        %cannot restore their value, resulting in memory corruption. 

        Two prerequisites must be satisfied to reliably execute Rowhammer: access to
        physically contiguous memory and fast, uncached memory access \cite{drammer, smash-rowhammer, dedup-est-machina, sledgehammer}. Physically contiguous memory enables two-sided Rowhammer which
        has been shown to be more effective at triggering bit flips \cite{anvil}. Fast, uncached memory access is required to activate aggressors in the same row at a sufficiently high rate to trigger bit flips.
        Modern DDR4 DIMMs implement Target Row Refresh or TRR to mitigate Rowhammer \cite{micron-trr}. While several undocumented, proprietary implementations of TRR exist, all of them essentially track memory accesses to identify aggressor rows and issue additional refreshes to the associated victim rows \cite{trrespass, utrr}.

    \subsection{Rowhammer for fingerprinting}
        \label{subsec:rh_as_a_fp}
        The rate at which a capacitor loses its charge depends on its
        physical properties~\cite{hennesyandpatterson, memorybook}. 
        These properties are not uniform on all chips due to process variation induced during manufacturing~\cite{SHOCKLEY196135}.
        This process variation also determines the susceptibility of each bit to flipping under Rowhammer.
        % As a result, the capacitors that lose their charge 
        % (i.e., the bits that flip) should depend on the particular chip being 
        % subjected to Rowhammer.
        Thus, when running Rowhammer with 
        identical parameters (under comparable environmental conditions) on 
        different chips, differences in the bit flip behavior can
        be attributed to differences in the physical properties of the DIMMs. 
        
        Rowhammer PUF \cite{rowhammer-puf} studies the distribution
        of bit flips on a PandaBoard \cite{pandaboard}, but they 
        do not compare the uniqueness of bit flips across devices. 
        They also present results on DDR2 memory which did not incorporate
        any mitigations against Rowhammer. Fingerprinters on a PandaBoard 
        also do not have to contend with the restrictions imposed by the OS 
        to access memory. Recently, researchers have extended the approach 
        proposed by Rowhammer PUF for fingerprinting \cite{fp-hammer} on 
        desktops with DDR4 DIMMs. However, they do not discuss ways to overcome 
        key challenges that complicate the extraction of fingerprints on 
        commodity devices:

        First, fingerprinters have to ensure that they compare bit flip distributions from the same memory chunk for fingerprinting. 
        However, fingerprinters have to rely on the abstractions provided by the 
        OS to access memory, which make it difficult to guarantee access to the 
        same chunk. Fingerprinters cannot identify devices by matching bit flip 
        distributions across different chunks of memory from the same module 
        since we find that each chunk has a unique distribution of bit flips 
        (Takeaway 1 in $\S$\ref{sec:measurement}). 
        
        Second, fingerprinters have 
        to overcome the non-deterministic behavior of bit flips 
        \cite{rowhammer-reproduction} when identifying devices. We see that bit 
        flips are non-deterministic since some capacitors are more susceptible 
        to Rowhammer (higher probability of flipping) when compared to others 
        (Takeaway 2 in $\S$\ref{sec:measurement}). This results in cases where 
        the approach proposed by prior research \cite{fp-hammer} is unreliable 
        in identifying devices even when building up references across multiple 
        hammering attempts (Figure \ref{fig:disjoint} and Figure 
        \ref{fig:union_flips}). Fingerprinters also have to contend with the 
        presence of external factors such as background applications and changes to the CPU's frequency (that they cannot control) which further impact the non-determinism in bit flips ($\S$\ref{sec:eval_freq}). 
        
        Third, the choice of the hammering pattern employed to overcome TRR also has an impact on which bits flip \cite{rowhammer-reproduction}. As a result, certain hammering patterns are more suitable for fingerprinting. However, prior research \cite{fp-hammer} does not discuss ways to identify patterns that help with fingerprinting.  Lastly, prior research presents results on a limited set of DRAM modules and does not discuss if their results generalize across DRAM modules of different configurations and manufacturers.

       With \system, we first make the observation that the bit flips in
       each contiguous 2 MB chunk of memory (one among several ways to allocate 
       contiguous memory using Transparent Huge Pages or THP \cite{thp}) are highly 
       unique and persistent ($\S$\ref{sec:measurement}). Leveraging this observation, 
       we propose a novel sampling strategy as part of \system's design 
       to overcome the memory restrictions imposed by the 
       operating system ($\S$\ref{sec:matching_phase}). When overcoming TRR to trigger bit flips for 
       fingerprinting, we operationalize hammering patterns at scale by 
       prioritizing those patterns that can trigger a large number of bit 
       flips ($\S$\ref{sec:templating_phase}). Such patterns help improve the
       robustness of the fingerprint in presence of external factors ($\S$\ref{sec:eval_freq}). To account for the inherent non-determinism in bit flips, we reset and hammer the memory 
       chunk multiple times to extract a probability distribution of bit flips ($\S$\ref{sec:hammering_phase}). We then compare the divergence of these probability distributions to reliably fingerprint DIMMs( $\S$\ref{app:eval_baseline}). Crucially, when compared to prior work, we always execute Rowhammer multiple times to ensure that our fingerprints are not affected by those capacitors that have a low probability of showing bit flips. We summarize how \system differs from prior research in $\S$\ref{sec:related_work}. \system is the first technique to demonstrate the extraction of unique and stable fingerprints on the largest scale using Rowhammer while overcoming practical limitations enforced by the OS and by Rowhammer mitigations such as TRR.
\vspace{-5mm}
\section{\system Overview}
\label{sec:overview}
%In this paper, we present \system, a Rowhammer-based fingerprinting approach that extracts unique and stable fingerprints even among devices that have identical hardware and software configurations. 
%In this section, we define our threat model and provide an overview of \system's architecture.

\subsection{Threat model}
In this paper, we take the role of the fingerprinter whose goal is to extract unique and stable fingerprints from devices even among those that have identical hardware and software configurations.
In our threat model, the fingerprinter can run code on a user's device. Specifically, we consider 
host-based fingerprinting \cite{clock-skew, rowhammer-puf, demicpu} where
users run a native application on their device that was developed
by the fingerprinter. We assume that the fingerprinter can
run unprivileged (without root privileges) 
native code on the user's device,
which attempts to extract a device fingerprint \cite{fpjs-android, fpjs-ios}.

As a malicious use case, multiple fingerprinters controlling two or more
different applications can use such fingerprints for cross-app
tracking \cite{reardon201950, aragorn}.  Malicious fingerprinters can use such fingerprints to execute
targeted attacks against specific victims \cite{clock-skew}. Fingerprinters cannot accomplish this using native identifiers such as IDFA \cite{idfa} or ADID \cite{adid} since they require user consent or can be reset by users. Fingerprinters could also use such fingerprints for security and
authentication \cite{demicpu}. For
example, game developers can use them to detect aim bots in 
multiplayer games \cite{hlisa} or detect and ban devices for cheating 
\cite{easy-anti-cheat}. 
%We discuss different use cases for host-based 
%fingerprinting in Appendix XX.

In this paper, we do not consider web-based fingerprinting (where a user 
visits a website controlled by the fingerprinter) since it is more challenging
to trigger Rowhammer from the browser \cite{sledgehammer}. We discuss these 
challenges and discuss ways to extend our approach to operate within the
browser in $\S$\ref{sec:web-extension}.

As mentioned in $\S$\ref{sec:background}, we focus on DIMM-based DRAM
packages in this paper, but our findings are also applicable to other
packaging techniques. We assume that the fingerprinter possess a wide array of devices and DRAM modules (DIMMs) with different configurations. Fingerprinters use these devices to discover ways 
to overcome mitigations such as TRR. 

\vspace{-3mm}
\subsection{\system architecture}
At a high level, \system triggers bit flips on multiple contiguous chunks of memory on a user's device and uses the distribution of
the triggered bit flips as a fingerprint. \system then uses a similarity
metric to compare fingerprints extracted from different sessions at different
points in time to recognize if these sessions were executed on the same 
device.

\system's operation consists of three phases, namely,  
a \emph{templating phase}, a \emph{hammering phase}, and a \emph{matching phase}. 

$\bullet$ In the templating phase, fingerprinters conduct experiments on their own devices to discover ways to overcome Rowhammer mitigations and trigger bit flips.

$\bullet$ In the hammering phase, fingerprinters execute code on users' devices. The 
fingerprinter's code uses the knowledge gained from the templating phase to 
trigger bit flips on their devices. They then create a probability 
distribution out of the triggered bit flips which serves as a fingerprint 
for the user's device.

$\bullet$ In the matching phase, fingerprinters compare the fingerprint extracted
from a user's device against other reference fingerprints to identify the
user. They also use the extracted fingerprints to create new or update
existing references.
\section{Bit Flip Measurement and Analysis}
\label{sec:measurement_all}
In this section, we first calculate a theoretical upper bound 
on the entropy that can be obtained from the bits that flip 
across multiple contiguous chunks of 2 MB of memory. While
calculating this theoretical upper bound, we assume that 
the process variation during the manufacturing of DIMMs is such that
the bits that flip within each row (on the same DIMM and across
DIMMs) are independent. In this analysis, we also assume that bit 
flips are deterministic, i.e., hammering the same memory regions
always results in the same set of bit flips. Then, we relax these
assumptions and validate our analysis by measuring the actual entropy 
on a set of 3,611 such chunks across 36 DIMMs. We focus on contiguous 
2 MB chunks of memory since we can obtain such chunks from the OS 
without requiring root privileges. Transparent Huge Pages is one of the methods employed by past Rowhammer 
research to allocate 2 MB chunks of contiguous memory. We highlight
that while we present \system in context of 2 MB huge pages, \system
can operate with any method of allocating contiguous memory \cite{sledgehammer, drammer}. From our measurement, we observe that bit flips are unique across
chunks and despite exhibiting non-deterministic behavior, are 
persistent within each chunk. We use takeaways from our measurement
study to design our fingerprinting technique.

\begin{figure}[h]
\centering
  \includegraphics[width=\columnwidth]{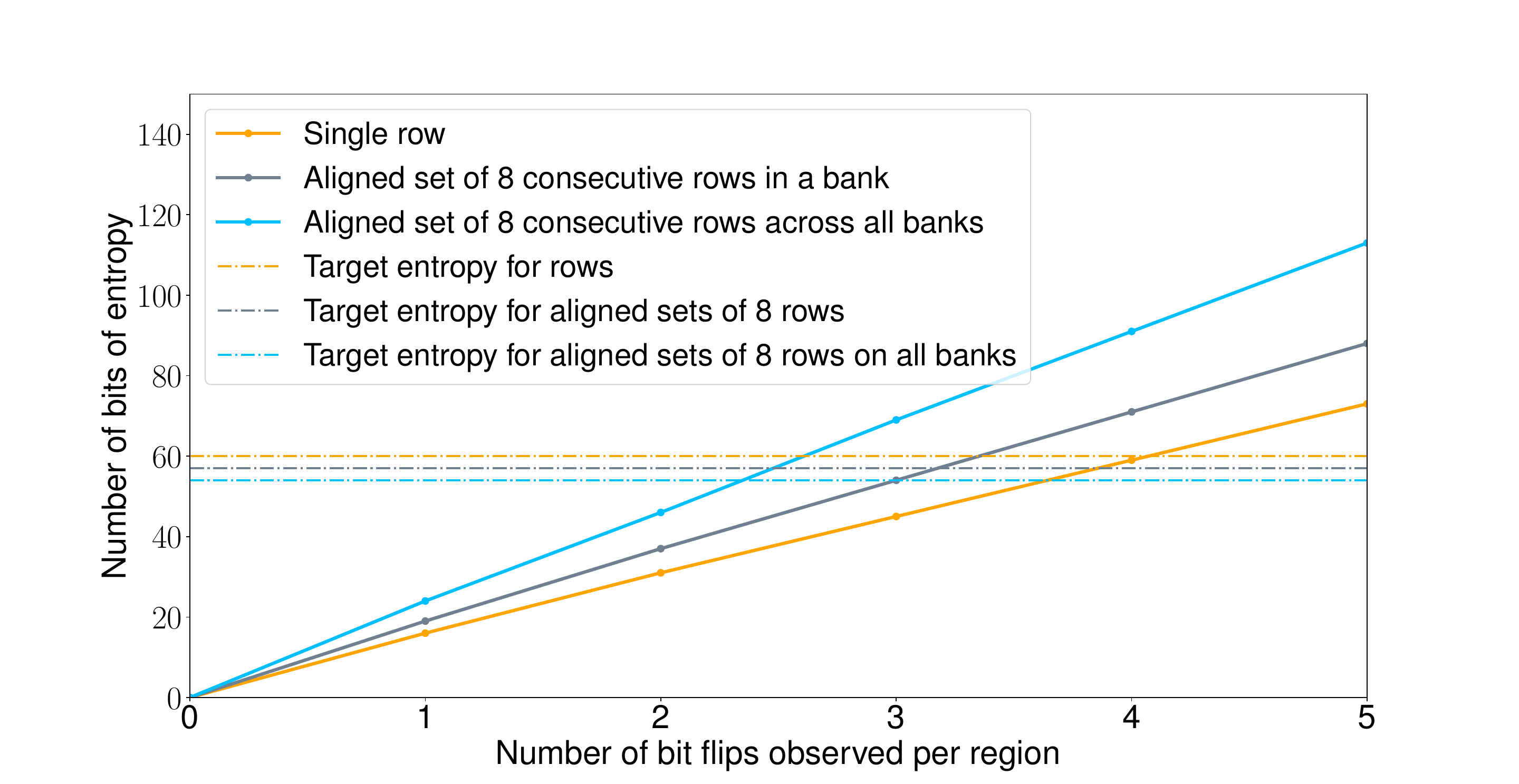}
  \vspace{-4mm}
  \caption{Plots showing the variation in the number of bits of entropy 
  that can be obtained to represent different regions of memory across
  a trillion DIMMs with varying number of bit flips.\\}
  \label{fig:theoretical_entropy}
\end{figure}
\vspace{-5mm}

\subsection{Theoretical entropy analysis}
If we consider that there are approximately 1 trillion DIMMs on the planet,
with each DIMM having approximately 10 banks and each bank having approximately
100,000 rows, we will have a total of $10^{18}$ possible rows. We will need
approximately $\log_2{(10^{18})} = 59.79 \approx$ 60 bits of entropy to represent all these rows.
Since rows within DRAM DIMMs are a finer granularity than the number of DIMMs 
(and correspondingly the number of devices), obtaining an entropy of 60 bits would
be sufficient to represent all devices on the planet.

Every row of a DRAM DIMM contains
65,536 capacitors. If process variation results in
Rowhammer triggering exactly one bit flip per row (i.e., only one capacitor losing 
its charge), we can use the index of the flipped bit within each row to at most 
identify 65,536 rows (equivalent of 16 bits of entropy). Thus, if exactly one bit
flips per row, using the index of the flipped bit within the row does not have enough
entropy to represent all possible rows across all DIMMs. The solid orange line in 
Figure \ref{fig:theoretical_entropy} shows the amount of entropy available to 
represent all rows with varying number of bit flips observed per row and the dashed 
orange line showing the required entropy. We see that if 5 bits 
flip per row, we can represent all possible rows since we get 73 bits 
of entropy ($\log_{2}{65,536 \choose 5}$).

The analysis presented so far limits us to only observe bit flips within a single row.
However, with a contiguous chunk of 2 MB of memory, we can access multiple rows from 
each bank. For example, in case of dual rank DIMMs having a width of 8 bits (going 
forward, we refer to this configuration as 2Rx8), a contiguous 2 MB chunk of memory 
corresponds to an aligned set of 8 consecutive rows (or 524,288 capacitors) within 
each bank. In this case, we will need an entropy of 57 bits to represent all $10^{17}$ 
chunks across all DIMMs. We see that if 4 bits flip among these rows, we get 71 bits of 
entropy ($\log_{2}{524,288 \choose 4}$) to represent them. As contiguous 2 MB chunks
are interleaved across banks, we can can access these consecutive rows across all
banks. If we consider all 16,777,216 capacitors spread across all banks in a 2 MB
chunk, we see that 3 bit flips in each chunk is sufficient to obtain an entropy of
54 bits ($\log_{2}{16,777,216 \choose 3}$). This entropy is sufficient to represent all 
$10^{16}$ such chunks across a trillion DIMMs. The grey and blue solid lines in Figure
\ref{fig:theoretical_entropy} show the variation in entropy with varying number of bit 
flips produced per aligned set of 8 consecutive rows and per aligned set of consecutive 
of 8 consecutive rows across all banks respectively. Dashed lines of the same colors 
show the required entropy in both cases.

In summary, our analysis indicates that the distribution of bit flips triggered 
by Rowhammer in individual 2 MB contiguous chunks of memory is potentially unique 
even if they can produce at least 5 bit flips. We reiterate that the 
theoretical analysis assumed that all chunks produce bit flips, the distribution 
of bit flips is independent and that bit flips do not exhibit any non-deterministic
behavior. We now perform experiments to measure the actual entropy across such 
chunks across multiple DIMMs.

\subsection{Empirical entropy analysis}
\label{sec:measurement}
Existing Rowhammer research has primarily focused on developing 
techniques to trigger bit flips~\cite{kim-rowhammer,rowhammer-js,drammer,trrespass,smash-rowhammer,blacksmith}
in memory. 
To the best of our knowledge, prior work lacks any analysis of the distribution of bit flips, particularly in terms of their entropy. 
In this section, we present the first such study on DDR4 DIMMs. Concretely, 
we first validate our theoretical analysis by measuring the entropy of the distribution 
of bit flips within a given bank across 
multiple 2 MB chunks of memory across DIMMs. As mentioned in $\S$\ref{sec:intro}, we find that bit flips
are not deterministic and, as a result, merely measuring the entropy of the distribution of bit flips is insufficient
to extract a reliable fingerprint. Thus, we also measure the persistence of the distribution of bit flips across 
repeated measurements to the same chunks across DIMMs.

\subsubsection{Test Bed}
Our test bed for this measurement consists of 36 identical
2Rx8 DIMMs. We fuzz one of these DIMMs to discover a non-uniform 
hammering pattern that can trigger bit flips. We observe that 
the discovered pattern triggers bit flips on all 36 DIMMs.

\subsubsection{Methodology}
We conduct our experiments in a controlled setting with
root privileges that allow us to allocate
1 GB of contiguous memory. This setting gives
us control over where we perform our hammering since we 
observe that the allocation of huge pages in physical memory
rarely changes in Linux (verified using pagemap~\cite{pagemap}).
We confine our hammering to randomly chosen contiguous 
2 MB chunks within the allocated huge page that lie within 
an arbitrarily chosen bank. While hammering, we modify the 
hammering pattern determined by the fuzzer to only trigger 
bit flips within the randomly chosen 2 MB chunks.

\begin{figure}[t]
\centering
  \includegraphics[width=\columnwidth,height=3.5cm]{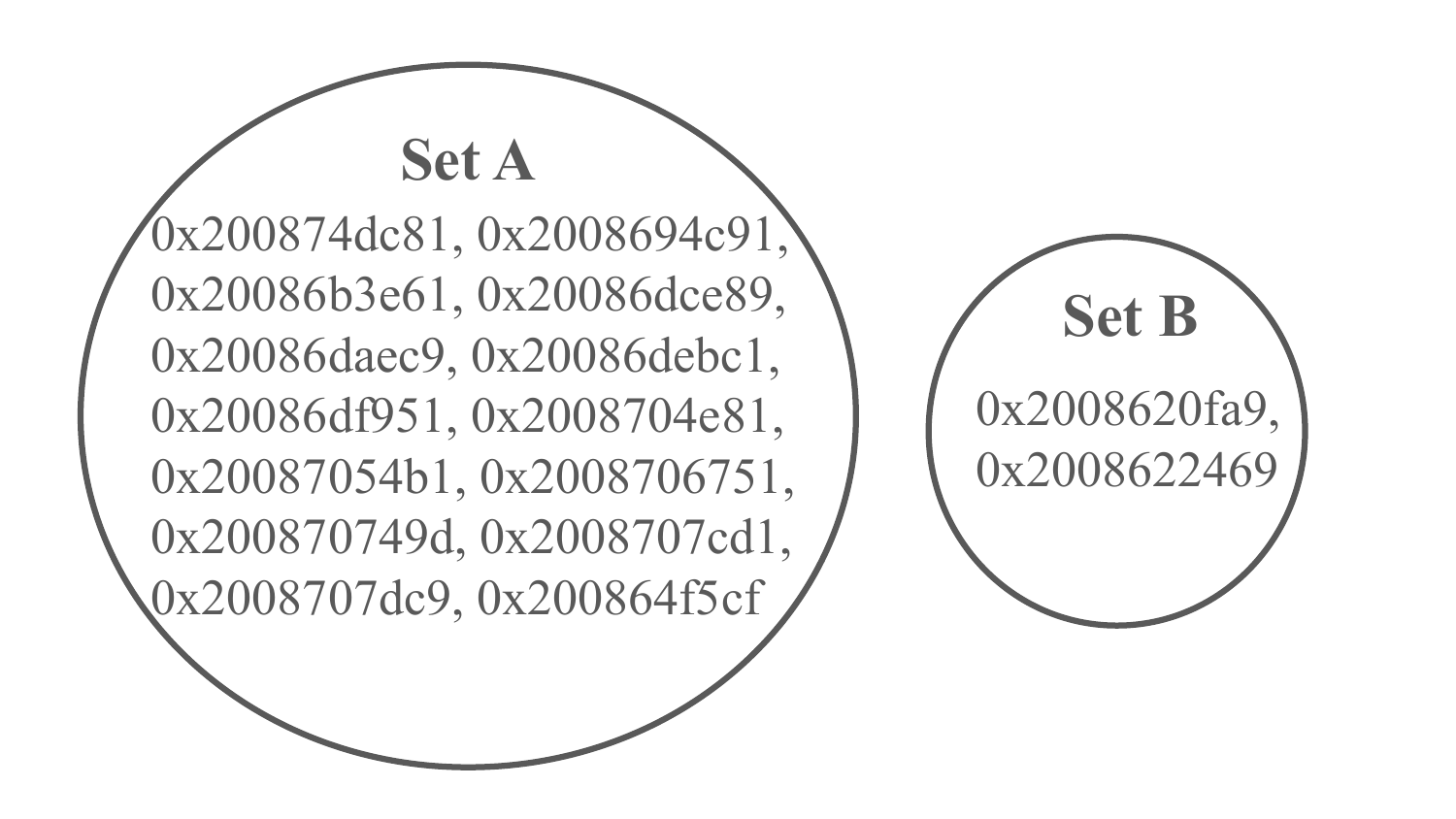}
  \vspace{-5mm}
  \caption{Set A shows the addresses of bits that flipped when hammering a particular chunk of a DIMM. Set B shows the addresses
  of bits that flipped when restoring data
  to the chunk and hammering it again. The two
  sets are disjoint demonstrating the
  non-deterministic behavior of bit flips.\\}
  \label{fig:disjoint}
\end{figure}

\begin{table}[b]
  \centering
  \small
  \begin{tabular}{|c|c|}
    \hline \hline
    {\textbf{Measurement}} & {\textbf{Value}}\\ \hline \hline
    Percentage of chunks showing bit flips & 99.77\%\\ \hline 
    Minimum number of bit flips per chunk & 1 flip\\ \hline
    Maximum number of bit flips per chunk & 1,799 flips\\ \hline
    Average number of bit flips per chunk & 711 flips\\ \hline
    Measured entropy across 3,603 chunks & 12 bits \\ \hline
    Normalized entropy & 1.0 \\ \hline \hline
  \end{tabular}
  \newline
  \caption{Summary of the measured distribution of bit flips across 36 identical DIMMs.}
  \label{fig:measurement_summary}
\end{table}

\begin{figure}[h]
\centering
  \includegraphics[width=0.8\columnwidth, height=3.5cm]{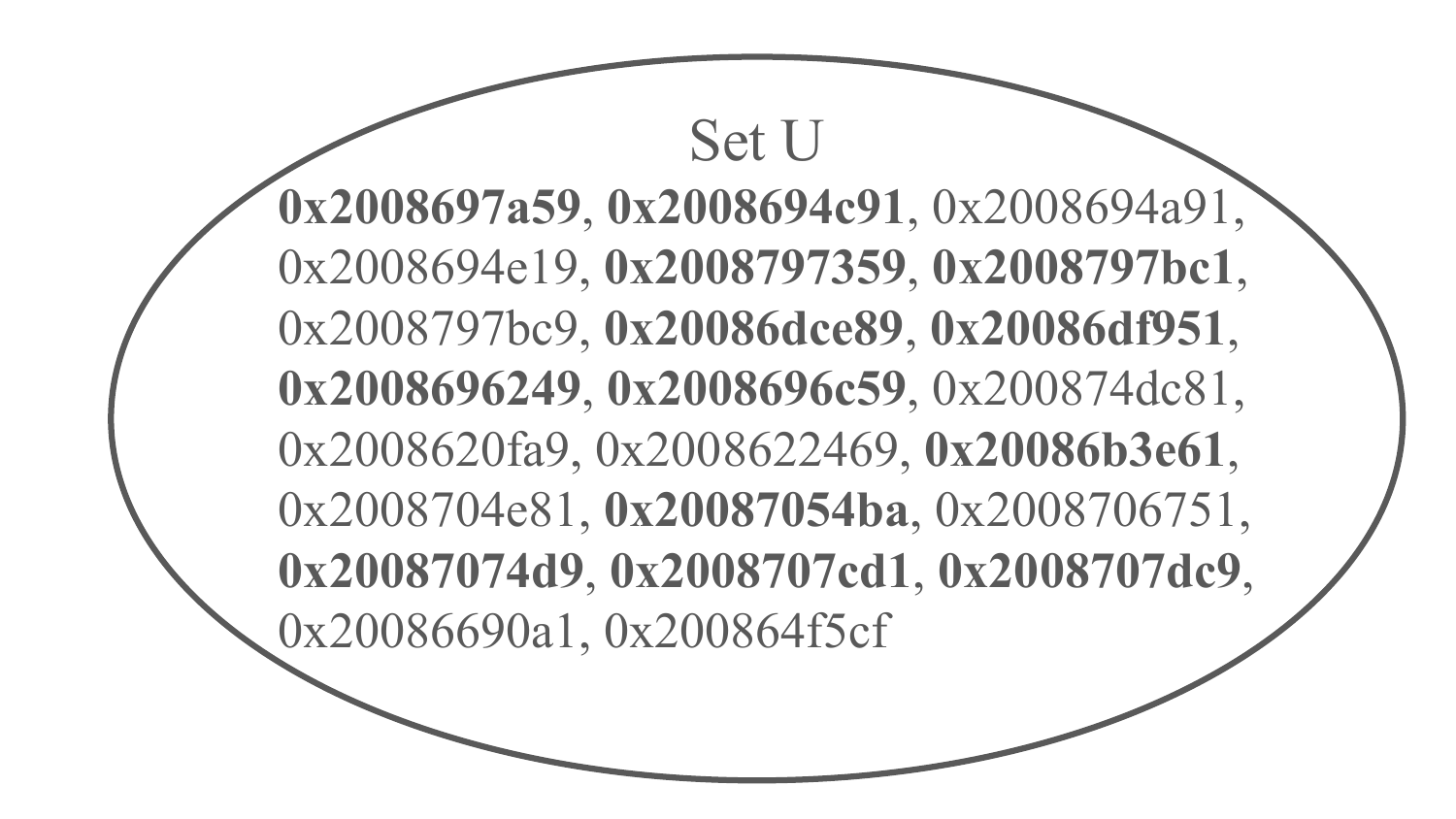}
  \vspace{-3mm}
  \caption{Set U shows the set of all addresses that flipped across 8 attempts to restore the data and hammering the same chunk from Figure \ref{fig:disjoint}. 
  The boldfaced addresses are addresses that flipped more than once across the 8 attempts. \\}
  \label{fig:union_flips}
\end{figure}
%\vspace{-5mm}
\subsubsection{Results}
Across all 36 DIMMs, we hammered a total of 3,611 chunks. 99.77\% of
these chunks (3,603 chunks) produced at least one bit flip. Among these
chunks, the number of bit flips ranged from 1 bit flip to 1,799 
bit flips at an average of 711 bit flips per chunk. Across the chunks
that produced bit flips, we record an entropy of 12 bits for the triggered
bit flips. We calculated entropy in terms of the number of chunks that had
the same set of bit flips as a given chunk. Crucially, we highlight that
in our test bed, 12 bits of entropy corresponds to the highest possible normalized entropy of 
1.0~\cite{norm-entropy}, which demonstrates that each chunk has a unique set of bit flips (i.e., all 3,611 chunks can be uniquely identified with 12 bits of entropy).  We summarize these findings in Table~\ref{fig:measurement_summary}. We use the fact that the bit flips in every chunk in our experiment is unique to
estimate the expected entropy on all possible chunks. Extrapolating
our results, based on the average of 711 bit flips per chunk yields over 7,700
bits of entropy, which is significantly higher than the 60 bits needed to represent
such chunks on a trillion DIMMs.
\newline\newline
\fbox{\begin{minipage}{\columnwidth}\textbf{Takeaway 1:} Every
contiguous 2 MB chunk of memory has a \emph{unique} set
of bit flips when
subjected to Rowhammer. 
\label{takeaway:chunk_unique}
\end{minipage}}
\newline

To use the bit flips produced by Rowhammer as a fingerprint, ensuring that they 
have high entropy on different chunks is not sufficient unless they are also 
persistent within the same chunk. In our study, we notice that reinitializing 
regions with the same data and hammering them again does not guarantee that the 
same bit will flip. In other words, we observe that the bits that flip within a 
given chunk are not deterministic.  For example, Figure~\ref{fig:disjoint} 
shows the sets of bits that flipped when hammering the same chunk on a particular 
DIMM twice (while restoring the data written to the chunk before hammering again).
Set A shows the addresses that flipped during the first attempt which is completely
disjoint to set B shows the set of addresses that flipped during the second attempt. 
%Thus, we cannot re-identify a particular chunk by merely employing a set similarity
%metric like Jaccard index (as proposed by prior research \cite{fp-hammer, rowhammer-puf}).
\newline\newline
\fbox{\begin{minipage}{\columnwidth}\textbf{Takeaway 2:} Bit flips
exhibit \emph{non-deterministic behavior}, i.e., hammering the same set of aggressor
rows multiple times does not result in the same set of bit flips. \end{minipage}}
\newline

\begin{figure}[!b]
\centering
  \includegraphics[width=0.9\columnwidth]{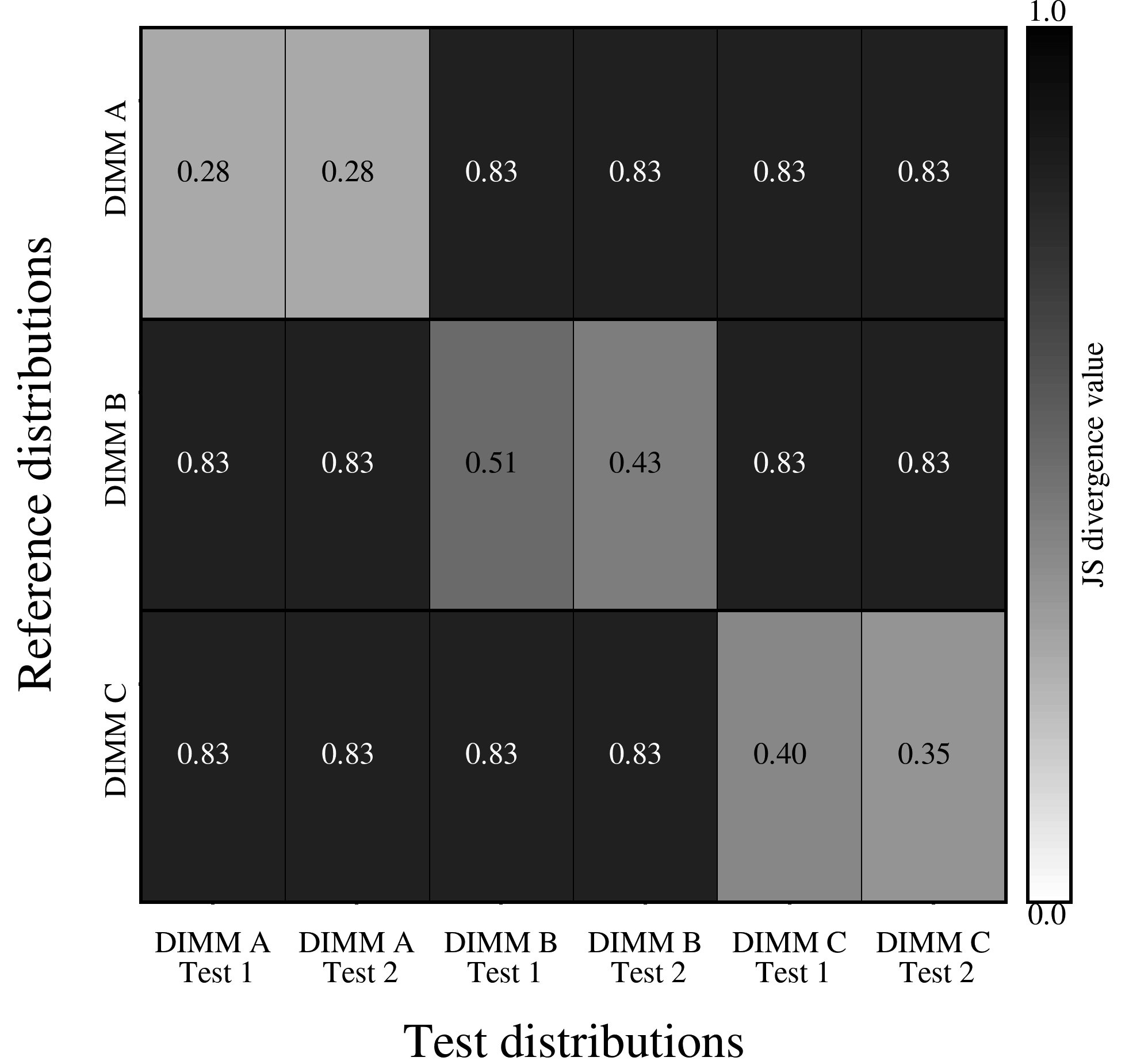}
  \vspace{-3mm}
  \caption{Visualization of the relative persistence of bit flips within given 2 MB chunks of memory across multiple DIMMs.\\}
  \label{fig:persistence_multiple}
\end{figure}

When we restored the data and attempted to hammer the same chunk 6 more times, 
we observed 2 attempts with no bit flips and 4 attempts where some addresses in
set A flipped again. We visually represent the list of bits that flipped across all 8 attempts in Figure \ref{fig:union_flips}. This figure shows that we cannot re-identify a particular chunk by merely employing a set similarity metric like Jaccard index
(as proposed by prior research \cite{fp-hammer, rowhammer-puf}). More importantly, the figure also indicates that some bits (such as those in set A)
have a higher probability of flipping and other bits (such as those in set B) have
a lower probability of flipping. Leveraging this observation, we compute a probability 
distribution for the bit flips in each chunk and match similarity of distributions to 
measure persistence. To extract a probability distribution from a given chunk, 
we hammer it multiple times and use the count of flips at different indices across 
all hammering attempts. 
Then, we use Jensen-Shannon (JS) divergence \cite{js-divergence} to compute the similarity of distributions. In Figure~\ref{fig:persistence_multiple}, we compare the 
distributions for 3 randomly chosen chunks from 3 different DIMMs across 3 different hammering attempts. 
We see similarities in the probability distributions computed from
the same chunk as compared to distributions across chunks.\newline
\fbox{\begin{minipage}{\columnwidth}\textbf{Takeaway 3:} Different bits
have different probabilities of flipping. The distribution of these probabilities
is \emph{unique} across contiguous 2 MB chunks of memory and \emph{persistent} within each 2 MB 
chunk. \end{minipage}}
%Thus, from our 
%measurement of the distribution of bit flips, we see that contiguous 2 MB chunks of 
%memory within a given bank has high entropy and high persistence.
\vspace{-2mm}
\section{\system}
\label{sec:design}
%We provide a detailed account of \system's three phases:
\subsection{Templating phase}
\label{sec:templating_phase}
In the templating phase, we (the fingerprinters) seek to discover hammering 
patterns that can overcome Rowhammer mitigations to trigger bit flips on our
own devices, so that we can employ the patterns to trigger bit flips on users' 
devices in the hammering phase. Concretely, in our experiments on DDR4 DIMMs (that employ in-DRAM TRR \cite{trrespass}), 
we run Blacksmith's \cite{blacksmith-code} fuzzer to discover non-uniform 
hammering patterns that can trigger bit flips. We observe that hammering 
patterns tend to be successful in evading TRR on multiple DIMMs from the 
same manufacturer. To account for all TRR implementations in the wild, our 
goal is to discover sufficiently many patterns to overcome all of them.

%To help speed up the discovery of patterns, we propose executing the fuzzer
%at the highest possible CPU frequency. We observed that executing a pattern
%(that is able to evade TRR) at a particular location when running the CPU 
%at a higher frequency produced more bit flips when compared to executing 
%the same pattern at the same location when running the CPU at a lower 
%frequency. From this observation, we hypothesize that even with a 
%pattern that can evade TRR, memory addresses in DRAM still have to 
%be accessed at a sufficiently high rate to produce bit flips.
%Thus, while running the fuzzer, we avoid discarding a pattern that 
%overcomes TRR but does not produce bit flips as a result of not being
%able to access addresses at the required rate by always operating at 
%the highest frequency. We note that we can set the CPU frequency 
%in this phase because we discover hammering patterns on our own 
%devices. When extracting fingerprints from users' device in the 
%hammering phase, we have to account for differences in bit flips 
%introduced by CPU frequency since we cannot set the frequency of 
%a user's device. We evaluate \system's robustness to changes in 
%CPU frequency in Section \ref{sec:eval_freq}.

Once the fuzzer discovers patterns that can trigger bit flips, we
evaluate them on our own DIMMs to decide which patterns to employ
on users' devices in the hammering phase. We prioritize those patterns 
that trigger more bit flips as well as those that generalize to 
trigger bit flips on more DIMMs.

Patterns that trigger more bit 
flips help account for differences in bit flip behavior when extracting 
fingerprints. These differences either arise as a result of the 
inherent non-deterministic behavior of bit flips ($\S$\ref{sec:measurement}) or as a result of external factors 
that fingerprinters cannot control. For example, 
we observe fewer bit flips (with the same pattern on the same DIMM) 
when running at lower CPU frequency ($\S$\ref{sec:eval_freq}). %Since we cannot control the 
%CPU frequency of a user's device, we pick patterns that can account 
%for changes to CPU frequency to ensure robustness of our fingerprints. 
A pattern that produces very few bit flips in the controlled setting 
of our own devices may not produce any bit flips on a user's device 
in presence of factors outside our control. In contrast, a pattern 
that produces a large number of bit flips in our controlled setting 
may still produce enough bit flips to be able to fingerprint a 
user's device. %We evaluate \system's robustness while extracting 
%fingerprints in context of varying CPU frequencies in $\S$\ref{sec:eval_freq}.

We observe that patterns that need fewer 
aggressors to overcome TRR (we refer to these as secondary aggressors) tend to trigger more bit flips. Intuitively, 
by virtue of being shorter patterns, these patterns would activate 
aggressors that trigger bit flips (we refer to these as primary
aggressors) at a higher frequency, resulting in more bit flips.
We prioritize patterns that generalize to trigger bit flips on more
DIMMs in our possession since they are also likely to generalize
to more DIMMs in the wild.
%In line with existing research \cite{blacksmith}, we observe that 
%patterns discovered on DIMMs from one manufacturer do not trigger 
%bit flips on DIMMs from a different manufacturer. 
%However, some 
%patterns trigger bit flips on more DIMMs from the same manufacturer when
%compared to other patterns.
\vspace{-3mm}
\subsection{Hammering phase}
\label{sec:hammering_phase}
In this phase, our goal is to trigger bit flips in a user's 
device and extract the distribution of bit flips as a fingerprint. Since we
do not have knowledge of the type of DIMMs (or their corresponding TRR implementations) 
on the user's device, we rely on the patterns discovered in the templating phase to 
trigger bit flips. The hammering patterns discovered by the fuzzer are defined by a 
set of aggressors, a phase, an amplitude and a frequency~\cite{blacksmith}.
The phase, amplitude and frequency are such that some aggressors engage with 
TRR (secondary aggressors) and the others trigger bit 
flips (primary aggressors). Within the execution of each pattern, we observe that the aggressors accessed at certain points in time always served as primary aggressors regardless of the choice of which addresses were used as primary and secondary aggressors. We also observe 
that the position of the primary aggressors within the pattern is fixed 
across all DIMMs where the pattern is able to trigger bit flips. Concretely, for a pattern $a_{1},a_{2},\ldots a_{i},a_{i+1},\ldots a_{n}$, addresses 
places at indices ${i}$ and ${i+1}$ served as primary aggressors on all
DIMMs where the pattern triggered bit flips. To reliably produce bit flips
in the hammering phase, we pick addresses such that the primary aggressors form a double-sided aggressor pair, and the secondary aggressors are other addresses
within the same bank as the primary aggressors. Fingerprinters can employ timing side channels (described in Appendix \ref{app:timing}) or other APIs (decode-dimms \cite{decode_dimms}, lspci \cite{lspci}, etc) to infer DRAM configuration/manufacturer and pick appropriate hammering patterns. 

\begin{figure}[t]
\centering
  \includegraphics[width=\columnwidth, height=5.5cm]{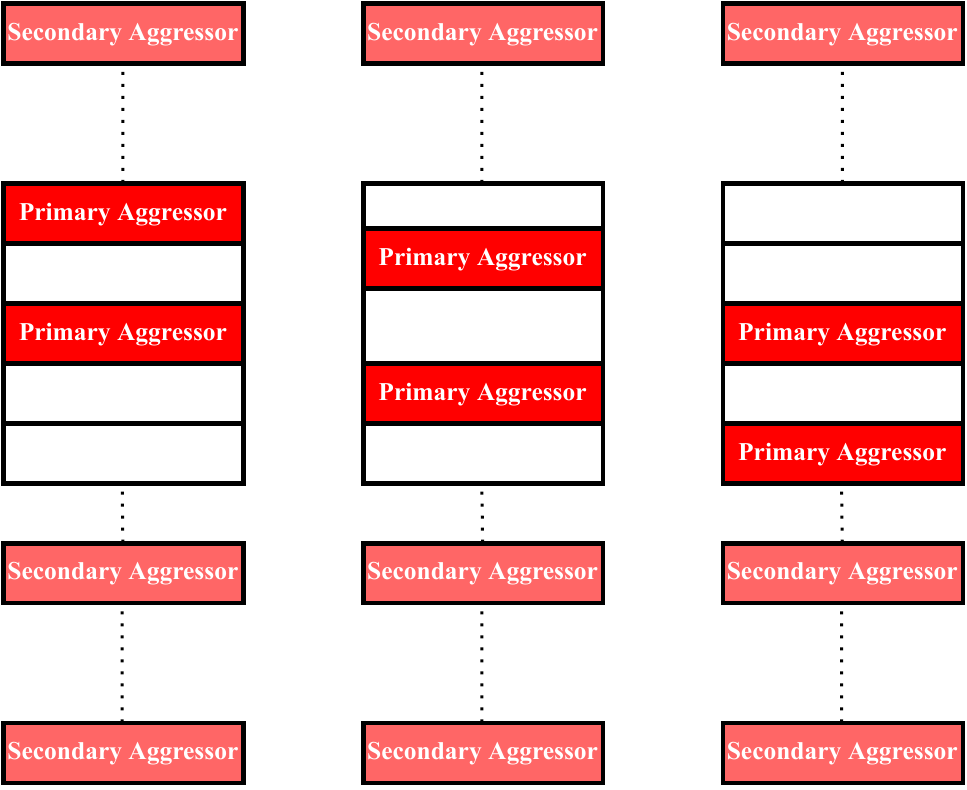}
  \vspace{-3mm}
  \caption{Visualization of a hammering sweep performed within one bank of a contiguous
  2 MB chunk. The central rectangular blocks in the visualization
  represent rows within the chunk. We map primary aggressors in the discovered
  patterns to rows within the chunk and secondary aggressors to
  random rows within the same bank. The hammering sweep involves sequentially
  hammering all pairs of double-sided aggressors as primary aggressors within the bank of the chunk  and scanning the other rows for bit flips. \\}
  \label{fig:hammering_sweep}
\end{figure}

In this section, we discuss the hammering phase in context of a single DIMM
present on the user's device. We discuss ways to extend the hammering phase
to devices having multiple DIMMs across multiple channels in Appendix \ref{sec:mult_chan}.
To execute the discovered patterns, we allocate transparent huge pages on a 
user's device to obtain contiguous chunks of 2 MB of memory without requiring root privileges. 
We can access 
all addresses within the huge page by modifying the lower 21 bits of the 
starting address of the chunk. This allows us to pick double-sided aggressor 
pairs since such chunks typically provide access to contiguous rows 
across multiple banks of a DIMM. For example, in case of the user's
device having one 1Rx8 DIMM, the chunk gives us access to 16 contiguous rows 
within each of the 16 banks on the DIMM. To trigger bit flips, we first 
choose a particular bank within the chunk. We can do this since most bits 
in the address that determine the bank are contained with the lower 
21 bits in most CPU architectures~\cite{drama,smash-rowhammer}. Then, we 
map the primary aggressors in the discovered patterns to double-sided 
aggressors within the chosen bank of the chunk and secondary aggressors to 
random addresses within the same bank (possibly outside the chunk). 
While we cannot determine the exact row of a given address from the lower 
21 bits, we do know the relative position of rows with respect to the row 
corresponding to the start address of the chunk. This allows us to choose 
different pairs of rows as primary aggressors within the chunk.

We sequentially consider all pairs of double-sided aggressors within the bank 
of the 2 MB chunk as primary aggressors and execute the discovered pattern. Upon 
executing the pattern with each pair of primary aggressors, we record the addresses 
that had bit flips within the allocated chunk and the corresponding change in the 
data written at those addresses. We then restore the original data written to the 
chunk, shift our primary aggressors to the next set of double-sided aggressors
within the chunk and repeat the same procedure. We refer to this operation 
of hammering all possible double-sided aggressors as primary aggressors 
within the allocated chunk as a hammering sweep. Figure \ref{fig:hammering_sweep} 
visualizes the \emph{hammering sweep}. 
To account for non-determinism 
in bit flips, we repeat the hammering sweep multiple times on the chunk. 
%Finally, we transmit the recorded bit flip information along with the starting address of the chunk to our server to match fingerprints.
\vspace{-3mm}
\subsection{Matching phase}
\label{sec:matching_phase}
With the observation from $\S$\ref{sec:measurement} that the distribution of 
bit flips within a bank of contiguous 2 MB chunks is highly unique and 
stable, we compare the similarity of these distributions to fingerprint 
them. From the information recorded in the hammering phase, we identify
the relative positions and counts of the capacitors that flipped within a 
contiguous 2 MB chunk (indexed from 0 to 1,048,576 in case of 1Rx8 DIMMs). 
We use these counts to create an empirical probability 
distribution for each capacitor to flip within the chunk. We then compare 
the similarity of this distribution against previously extracted distributions
using JS divergence to identify the chunk. In case the newly extracted distribution is significantly different from all previously extracted distributions, we consider the distribution to have been extracted from a new DIMM and use the distribution as a reference for that DIMM.

However, we cannot guarantee access to the same 2 MB chunks on a user's 
device, since memory allocation is handled by the OS. Thus, if we
obtain two different 2 MB chunks of memory on a user's device during two
different sessions and compared their distributions, we would incorrectly 
conclude that different devices were used during these sessions. For example, the OS may allocate a particular huge page in one session and a different huge page from the same DIMM in a subsequent session. The bit flip probability distribution extracted from these huge pages would be different
since the distribution is unique for different 2 MB chunks.

We overcome this challenge by taking inspiration from the birthday paradox 
\cite{abramson1970more, dedup-est-machina}, which shows that there is a high
probability for at least one entity to be present in multiple groups even for
modest-sized groups. Concretely, we hammer multiple 2 MB chunks during each
session to ensure that we hammer at least one previously hammered chunk. For example, suppose the user's device has 1 GB of memory which corresponds
to 512 different 2 MB chunks of memory. From the birthday paradox, if we were to hammer 64 chunks each in two different sessions with the user's device, 
then the probability that at least one chunk would overlap between them is 
over $99.9\%$. We derive this from first principles in Appendix \ref{app:math}.

One drawback to this approach is that hammering a large number of chunks would 
prolong the duration of the hammering phase, thereby making it less efficient.
However, we can overcome this by building up references across 
sessions, which would result in hammering fewer chunks in the long 
term. For example, suppose we have reference distributions to 64 different 
chunks from one session. In a subsequent session with the same device, we hammer 64 
chunks such that only one chunk happens to overlap with the 
reference. We can now combine the distributions of the 63 non-matching chunks to our reference to have an updated reference from 127 different chunks from that
device. When running a subsequent session on the same device, we have a higher 
probability that an allocated chunk would match our reference, since the reference 
size has increased. Upon reaching the limiting case where we have references for all possible chunks, merely hammering 
one chunk in the hammering phase would be sufficient, thereby 
resulting in higher efficiency. We show the progressive decrease in number of chunks to sample with growing reference sizes in Appendix \ref{app:math}.
\section{Evaluation}
\label{sec:eval}
We evaluate \system in terms of the
uniqueness of its fingerprints, stability of its
fingerprints, time taken to extract fingerprints and
robustness of its fingerprints in presence of 
external factors. We also compare \system's approach
of matching probability distributions against the 
approach of matching Jaccard similarity.

    \begin{figure*}[tb]
        \begin{center}
            \mbox{ 
            %\hspace{-1.0ex}
            \subfigure[Distribution of JS divergence values across all pairs of 
            fingerprints from 36 identical 2Rx8 DIMMs]
                {
                \label{fig:short_2rx8}
                \includegraphics[width=0.3\textwidth]{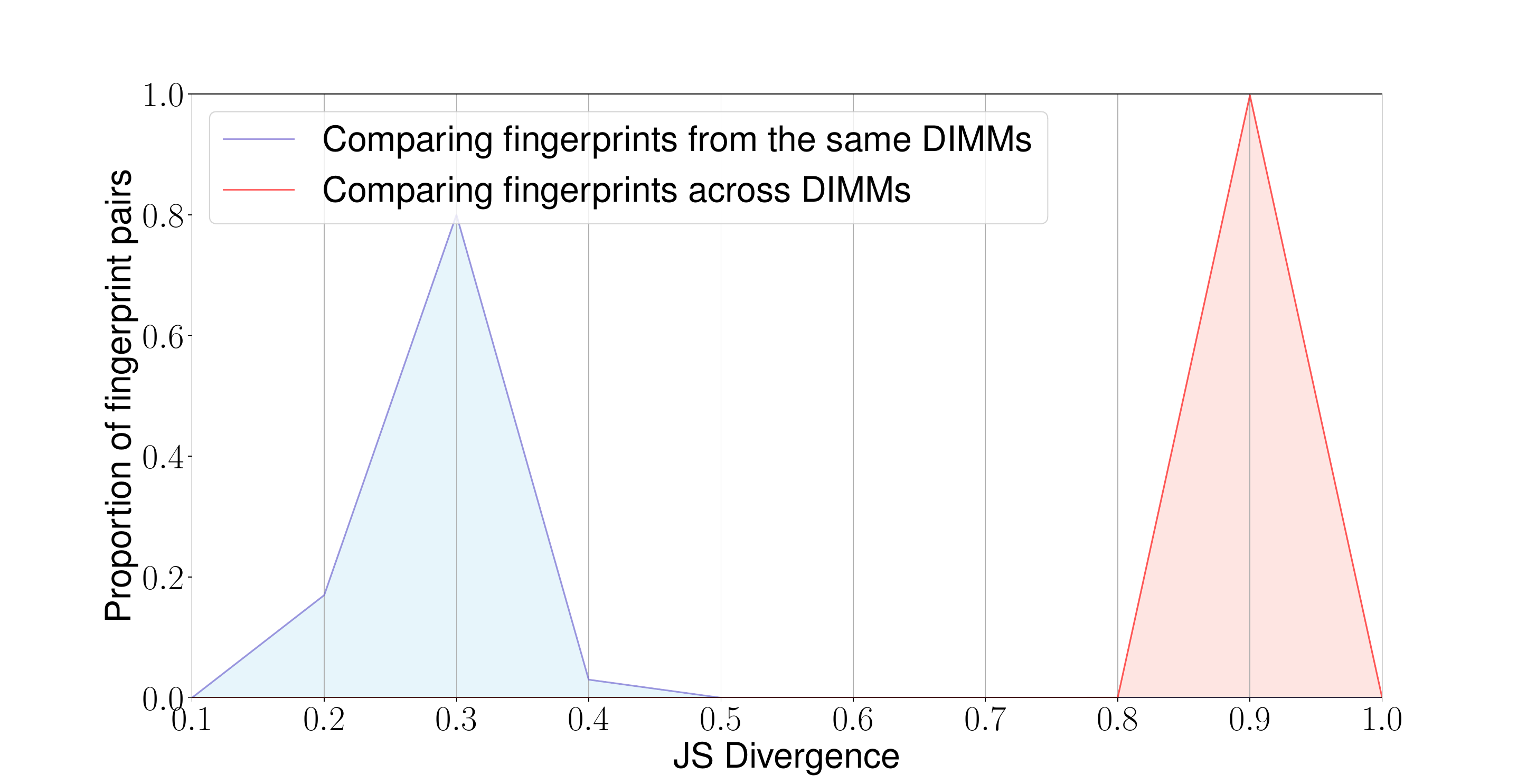}
                }
            %\hspace{1.0ex}
            \subfigure[Distribution of JS divergence values across all pairs of 
            fingerprints from 35 identical 1Rx8 DIMMs]
                {
                \label{fig:short_1rx8}
                \includegraphics[width=0.3\textwidth]{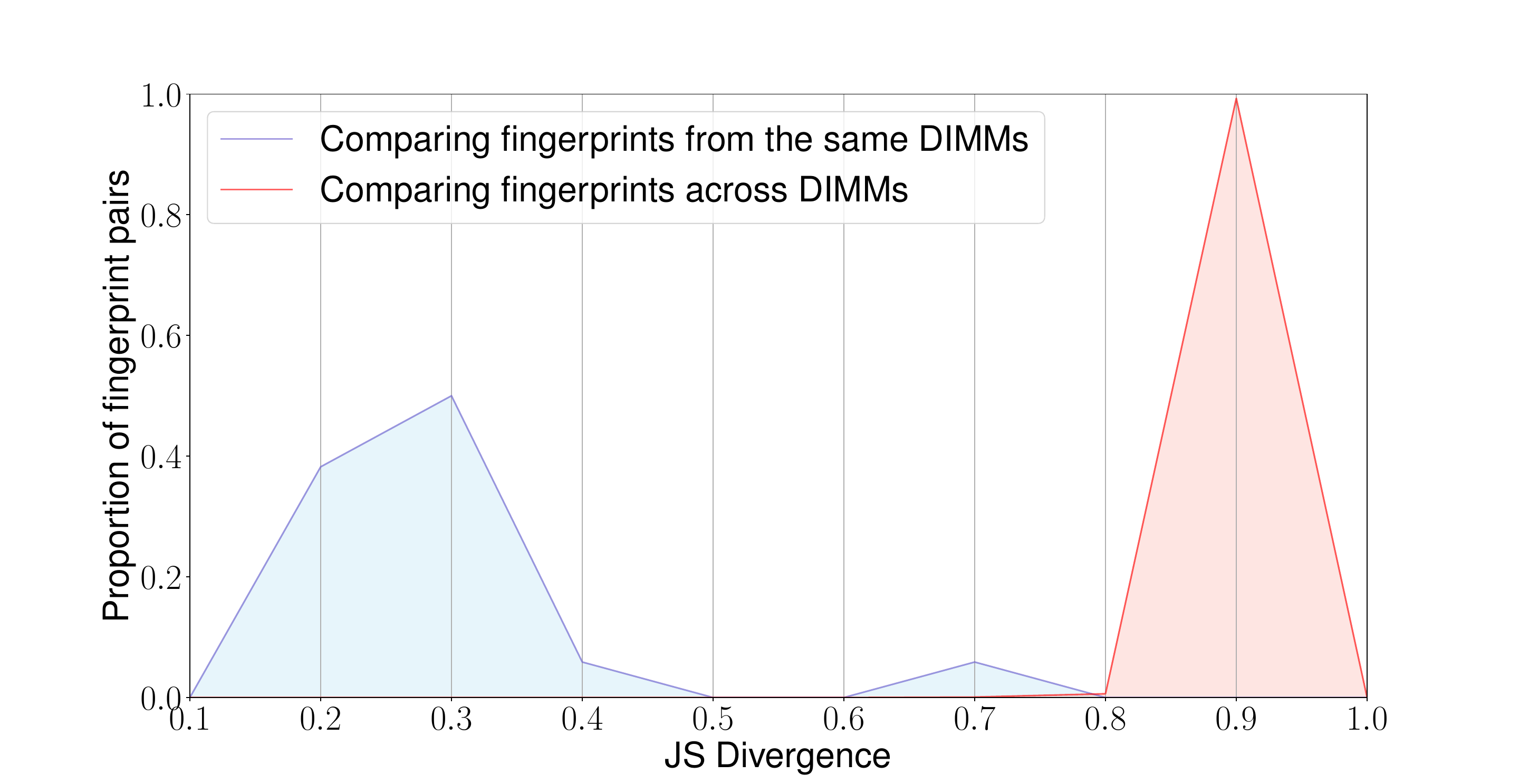}
                }
            %\hspace{1.0ex}
            \subfigure[Distribution of JS divergence values across all pairs of 
            fingerprints from 11 identical 1Rx16 DIMMs]
                {
                \label{fig:short_1rx16}
                \includegraphics[width=0.3\textwidth]{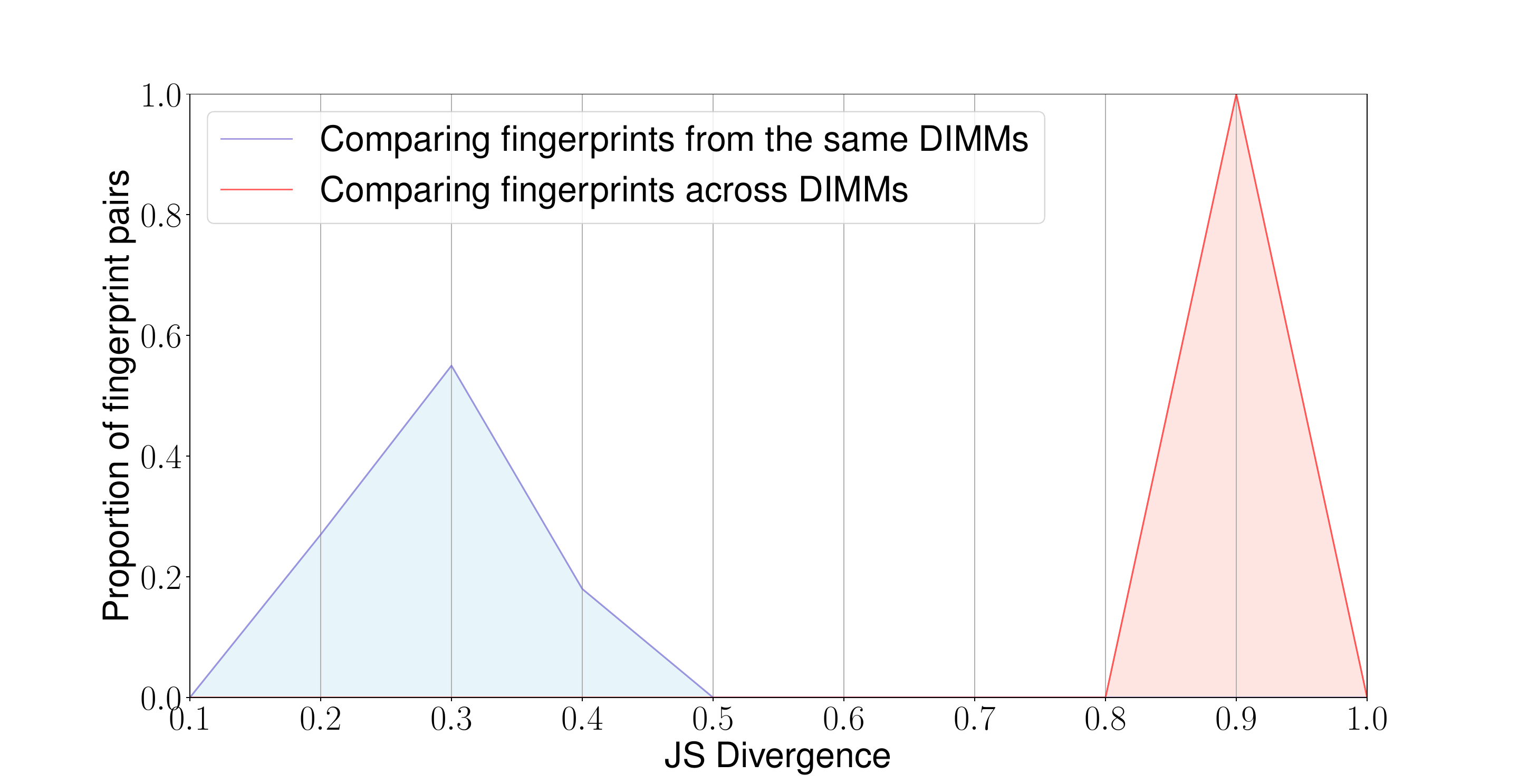}
                }
            }
            \vspace{-3mm}
            \caption{Plots showing the distribution of JS divergence values when
            comparing bit flip distributions obtained from the same pair of DIMMs and
            across different DIMMs.\\}
        \end{center}
    \end{figure*}

\subsection{Test bed}
Since the likelihood of an unintended bit flip that results in a crash is non-trivial, we do not evaluate \system in the wild. Instead, we evaluate \system on
a test bed in our lab.

Our test bed consists of 35 single rank DIMMs having a width of
8 bits (1Rx8), 11 single rank DIMMs having a width of 16 bits (1Rx8)
and 36 dual rank DIMMs having a width of 8 bits (2Rx8) from one
major DRAM manufacturer. Our test bed also contains 10 1Rx8 DIMMs,
2 1Rx16 DIMMs and 4 2Rx8 DIMMs from another major DRAM manufacturer. 
Overall, our test bed contains 98 DIMMs that include 6 indentical sets 
across 2 major manufacturers.
%Overall, our test bed contains 98 DIMMs that include 6 identical sets
%of DIMMs across two major DRAM manufacturers. 
We installed these DIMMs among 12 Intel Core i7 desktops (2 Skylakes, 9 Kaby Lakes and one Coffee Lake) for our experiments. %Since we have more DIMMs than desktops in our possession, we hammer DIMMs in batches. 
While repeating experiments on a particular DIMM, we make sure that we seat it on the same slot on the same desktop. While we focus on Intel desktops running Linux in our experiments, \system is not
specifically tied to a particular processor or operating system\footnote{Researchers have already demonstrated Rowhammer on other processors \cite{drammer, zenhammer} and operating systems \cite{drammer, windows-rowhammer}}. Bit flip data from our test bed
is available at \href{https://osf.io/cyp74/?view_only=3365b27d52464606a1a455014396a1c2} {this link} to help future research on Rowhammer.

\subsection{Experimental methodology}
We use a set of 11 hammering patterns, which together,
trigger bit flips on all 98 DIMMs in our test bed. We note that patterns generalize across DIMMs that were made by the same manufacturer. In other words, patterns that produce bit flips on DIMMs of a particular manufacturer do not produce bit flips on DIMMs from other manufacturers. Thus, the pattern that 
triggers bit flips on a DIMM also identifies
its manufacturer.

For each DIMM in our test set, we perform 
a hammering sweep (visualized in Figure~\ref{fig:hammering_sweep}) on a particular 
bank of multiple 2 MB 
chunks of memory with the appropriate pattern and record the resulting distribution 
of bit flips in that chunk. To compute the resulting distribution, we repeat
the hammering sweep operation multiple times on each chunk. 
 We consider the set of bit flip probability distributions of all hammered chunks 
 on a DIMM as its fingerprint. %We vary the number of times we repeat 
% the hammering chunk operation and the number of times we activate aggressors
%in different experiments to evaluate their impact on \system's fingerprints ($\S$\ref{sec:eval_eff}).

Given two fingerprints, we compute the JS divergence
on all pairs of bit flip probability distributions between them. We consider the two
fingerprints to match (i.e., to correspond to the same DIMM) if the minimum JS 
divergence value across all pairs is below an empirically determined threshold.
\vspace{-2mm}
\subsection{How unique are the fingerprints extracted by \system?}
\label{sec:eval_uniq}
For this evaluation, we extract 2 fingerprints from each DIMM in our test bed using 
the aforementioned methodology. We use the first fingerprint extracted
from a particular DIMM as a reference for that DIMM and compare subsequently extracted
fingerprints against the reference. In these experiments, we extracted both fingerprints 
within the space of few hours on the same day. We did not re-seat the DIMMs in the interim
period. We repeated the hammering sweep operation 8 times and activated aggressors
10,000,000 times.

Figure~\ref{fig:short_2rx8} shows the recorded minimum JS divergence when matching fingerprints from
each of the 36 2Rx8 DIMMs against reference fingerprints 
from each of them. From the figure, we clearly see that the minimum JS divergence 
computed among distributions taken from the same DIMMs is significantly lower than 
the minimum JS divergence computed among distributions taken across different DIMMs.
Figure~\ref{fig:short_1rx8} and Figure~\ref{fig:short_1rx16} shows similar plots across 35 1Rx8 DIMMs and 11 1Rx16 DIMMs respectively. We report similar plots for the other manufacturer in Appendix \ref{app:alt_manuf}.

The clear separation in JS divergence computed on pairs of fingerprints (bit flip distributions) taken from the same DIMM against those taken from different DIMMs 
allows us to pick multiple thresholds for JS divergence to uniquely identify DIMMs.
By picking appropriate thresholds\footnote{We picked thresholds by running the same experiment on a smaller subset of DIMMs.}, we attain an overall fingerprint accuracy of 
99.91\%, corresponding to a precision of 100\% and recall of 97.06\%. 
We highlight that even though we incrementally grew the size of our test bed 
by purchasing new DIMMs, we did not have to tune thresholds to attain the reported accuracy. Overall, these results show that \system has very high discriminative power and can uniquely identify DIMMs across multiple sets of DIMMs with identical configurations.

\begin{comment}

\begin{table}
  \centering
  \small
  \begin{tabular}{|c|c|c|c|}
    \hline \hline
  {\textbf{Number of DIMMs}} & {\textbf{Accuracy}} & {\textbf{Precision}} & {\textbf{Recall}} \\ \hline \hline
  {98} & {99.91\%} & {100\%} & {97.06\%}\\ \hline \hline
  \end{tabular}
  \newline
  \caption{\system's precision, recall and accuracy when matching fingerprints across 
  different sets of identical DIMMs.}
  \label{fig:eval_uniqueness}
  \hrulefill
\end{table}

\end{comment}

\subsection{How stable are the fingerprints extracted by \system?}
We evaluate the stability of the fingerprints extracted by
\system over time. Since we have more DIMMs than desktops, we first evaluate
\system's stability on a set of 10 random DIMMs across both manufacturers (6 and 4 DIMMs
respectively) by extracting fingerprints from them once a day for 10 days. We 
highlight that we do not re-seat DIMMs at anytime during this evaluation which
is what we would expect from users in wild. We repeated the hammering sweep operation 
8 times and activated the aggressors 200,000 times.

\begin{figure}[h]
\centering
  \includegraphics[width=0.98\columnwidth]{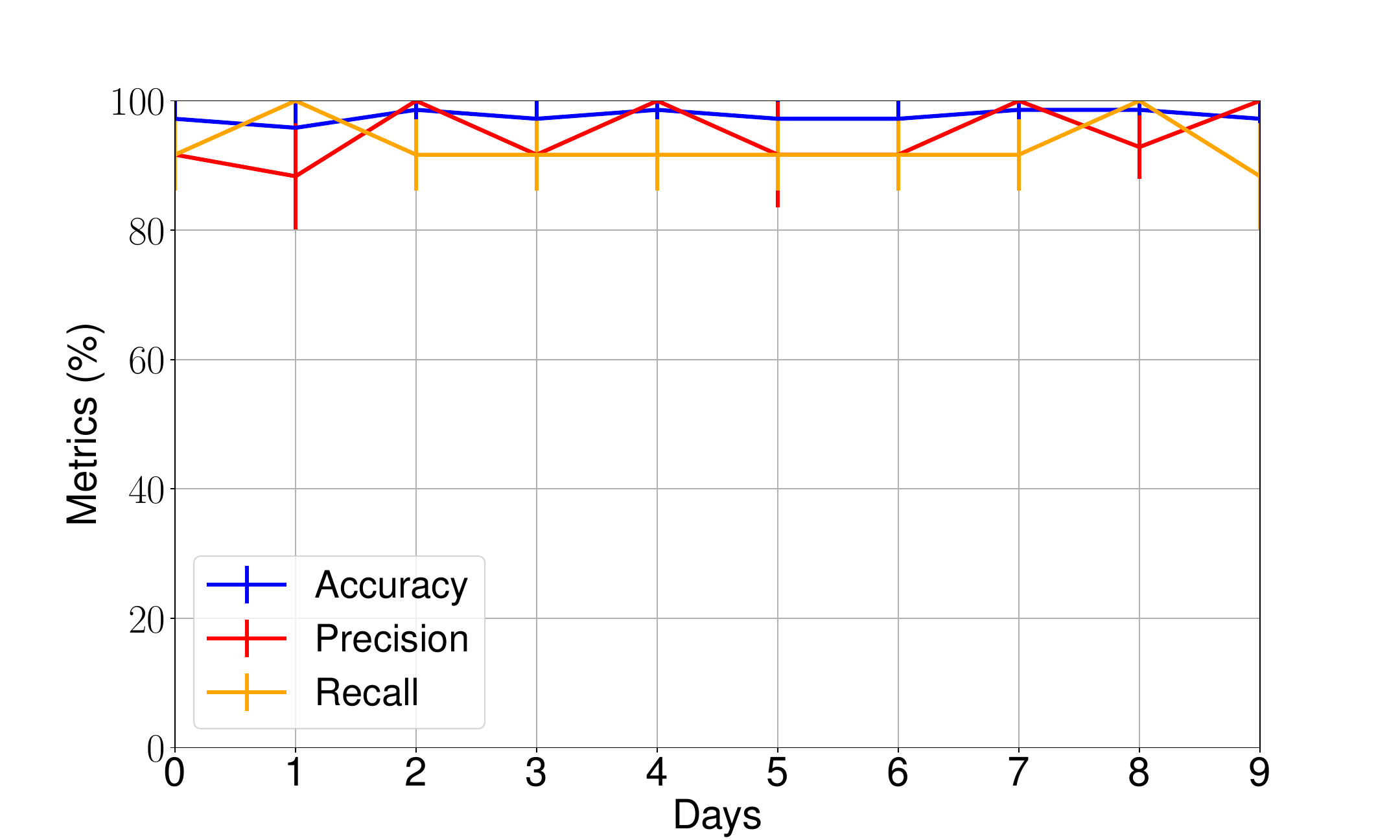}
  \vspace{-3mm}
  \caption{Variation in the accuracy, precision and recall
  of the fingerprints extracted by \system on a set of 10 DIMMs over 
  10 days. The metrics roughly remain constant with minor 
  fluctuations which provides strong evidence that the fingerprints extracted by \system are
  stable.}
  \label{fig:short_stability}
\end{figure}

Figure \ref{fig:short_stability} shows the variation in \system's accuracy, precision 
and recall on these DIMMs over time. From the plot we see that all metrics roughly remain 
constant with some minor fluctuations. Importantly, the plot does not show any trend of 
decline which indicates that the fingerprints extracted by \system 
are stable. These metrics were computed using the same threshold for each day which 
indicates that the JS divergence values (and the corresponding bit flip probabilities)
remain unchanged. We highlight that the stability is not a result of our specific choice
for the threshold since we record similar accuracy, precision and recall even with slightly
altered thresholds.
\begin{figure}[h]
\centering
  \includegraphics[width=0.98\columnwidth]{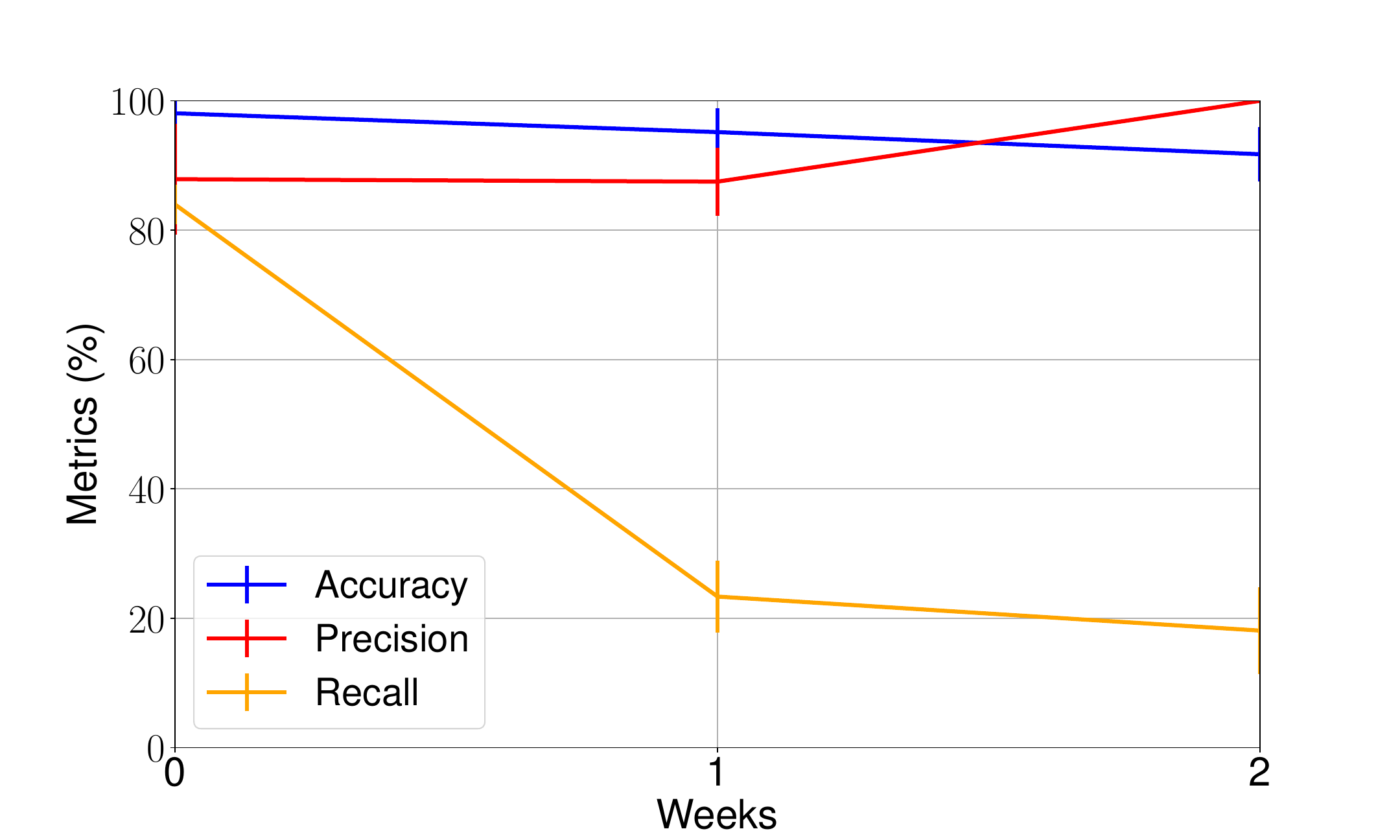}
  \vspace{-3mm}
  \caption{Variation in the accuracy, precision and recall
  of \system's fingerprints extracted on our entire test bed (98 DIMMs) 
  over 2 weeks and again over 4 weeks. Unlike our previous observation, 
  there is a change in metrics with the recall declining significantly. We 
  suspect that this decline emerges from having to re-seat DIMMs in our 
  experiments.\\}
  \label{fig:long_stability}
\end{figure}

Motivated by these results, we increased the scale of our evaluation by evaluating on
all DIMMs, first over a period of 2 weeks and then over a period of 4 weeks. For these
experiments, we had to re-seat DIMMs in the interim period to cover all DIMMs due to
the limited number of desktops at our disposal. Figure \ref{fig:long_stability} shows 
how \system's accuracy, precision and recall change over two weeks and over four 
weeks on our entire set of DIMMs. At a common threshold of 0.8 for JS divergence, 
we only see minor fluctuations in precision and accuracy, but we see a significant 
decline in recall. When we tried to change the threshold to improve the recall, 
we observed that it came at the cost of precision. We were unable to pick a common 
threshold that gave us high precision and high recall. 

We suspect that this instability in \system's fingerpints is a result of our
experimental setup where we re-seat DIMMs. If this is indeed the case, \system's
fingerprints would be stable in the wild since users rarely re-seat their DIMMs.
The ideal way to confirm our suspicion would be to evaluate the stability of \system's 
fingerprints on our entire test bed over an extended period of time (4 weeks) 
without having to re-seat DIMMs. Since this is not feasible due to the limited
number of desktops at our disposal, we run experiments that provide strong
evidence that re-seating induces the instability in \system's fingerprints. 
\begin{table}
  \centering
  \small
  \begin{tabular}{|c|c|c|c|}
    \hline 
  {\textbf{Experiment}} & {\textbf{Accuracy}} & {\textbf{Precision}} & {\textbf{Recall}} \\ \hline \hline
  {No rebooting/re-seating} & {100\%} & {100\%} & {100\%} \\ \hline
  {Only Rebooting} & {100\%} & {100\%} & {100\%} \\ \hline
  {Re-seating (and rebooting)} & {91.67\%} & {95.83\%} & {50\%} \\ \hline 
  \end{tabular}
  \newline
  \caption{Even when comparing fingerprints extracted over a span of a few minutes, we see
  that there is a significant drop in recall only when we re-seat DIMMs.\\}
  \label{fig:eval_stability}
\end{table}

Concretely, we run three different experiments on six randomly chosen identical DIMMs from our
test bed. In the first experiment, we hammer and extract two fingerprints within the
space of a few minutes. We also extract two fingerprints within the space of a 
few minutes in the second experiment, but we re-seat the DIMM (i.e., take out
and insert back) after extracting the first fingerprint. We do the same thing
in the last experiment but reboot the desktop after extracting the first 
fingerprint. Since we cannot re-seat a DIMM without rebooting the device,
we run the third experiment to see the impact of rebooting on the stability
of our fingerprint.

\begin{figure*}[tb]
        \begin{center}
            \mbox{ 
            \hspace{-1.0ex}
            \subfigure[Plots when repeating the hammering sweep operation 8 times]
                {
                \label{fig:tradeoff_8}
                \includegraphics[width=0.3\textwidth]{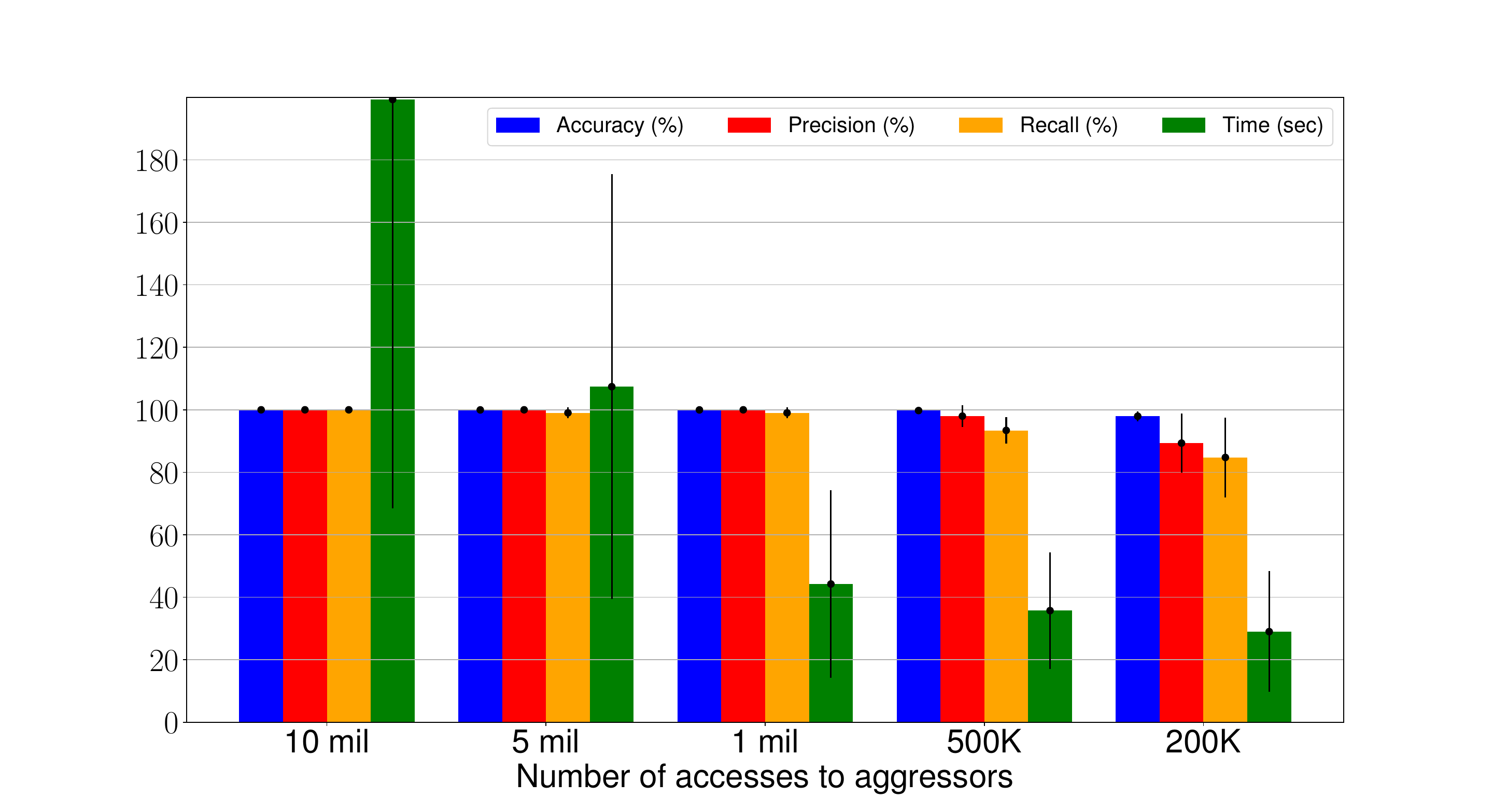}
                }
            \hspace{1.0ex}
            \subfigure[Plots when repeating the hammering sweep operation 4 times]
                {
                \label{fig:tradeiff_4}
                \includegraphics[width=0.3\textwidth]{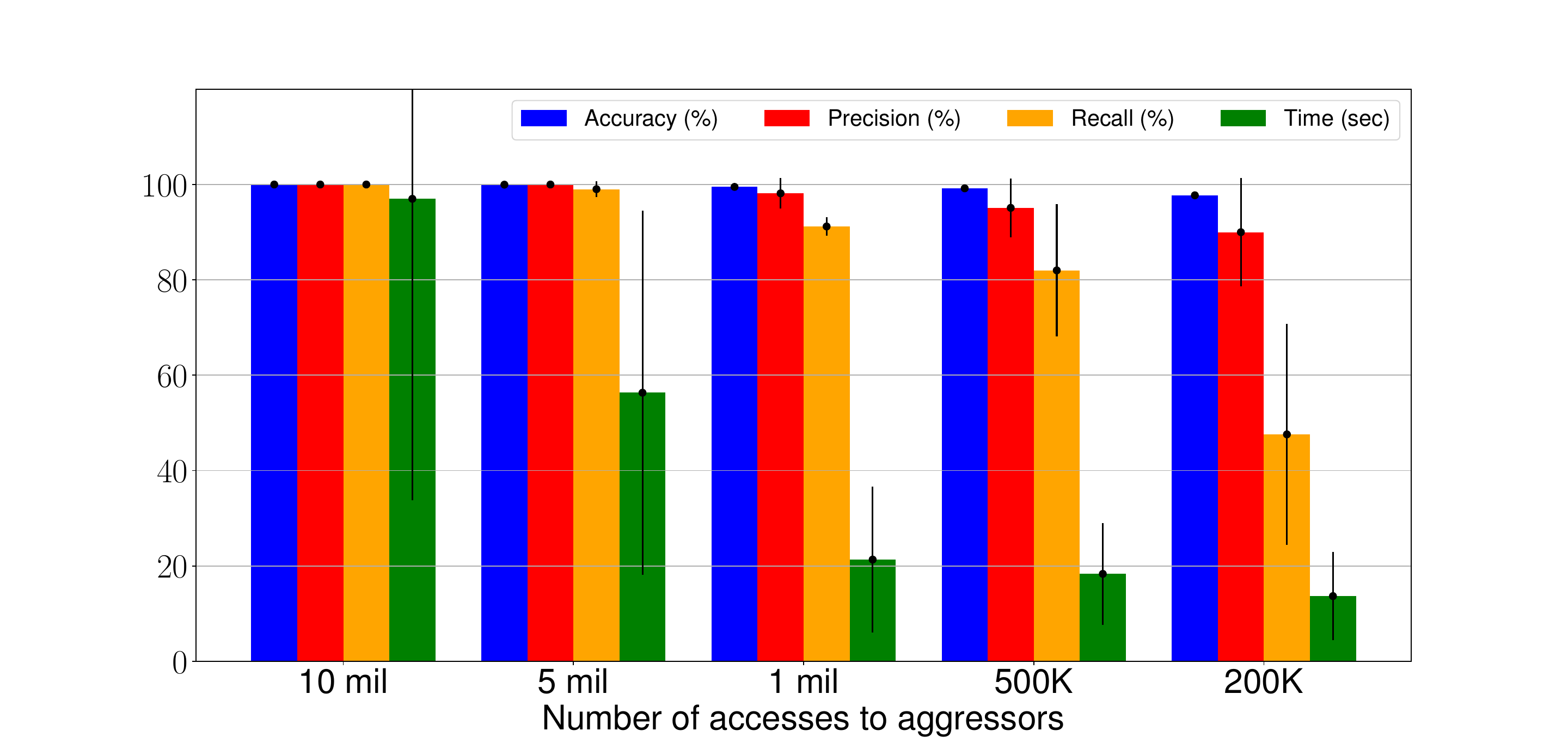}
                }
            \hspace{1.0ex}
            \subfigure[Plots when repeating the hammering sweep operation 2 times]
                {
                \label{fig:tradeoff_2}
                \includegraphics[width=0.3\textwidth]{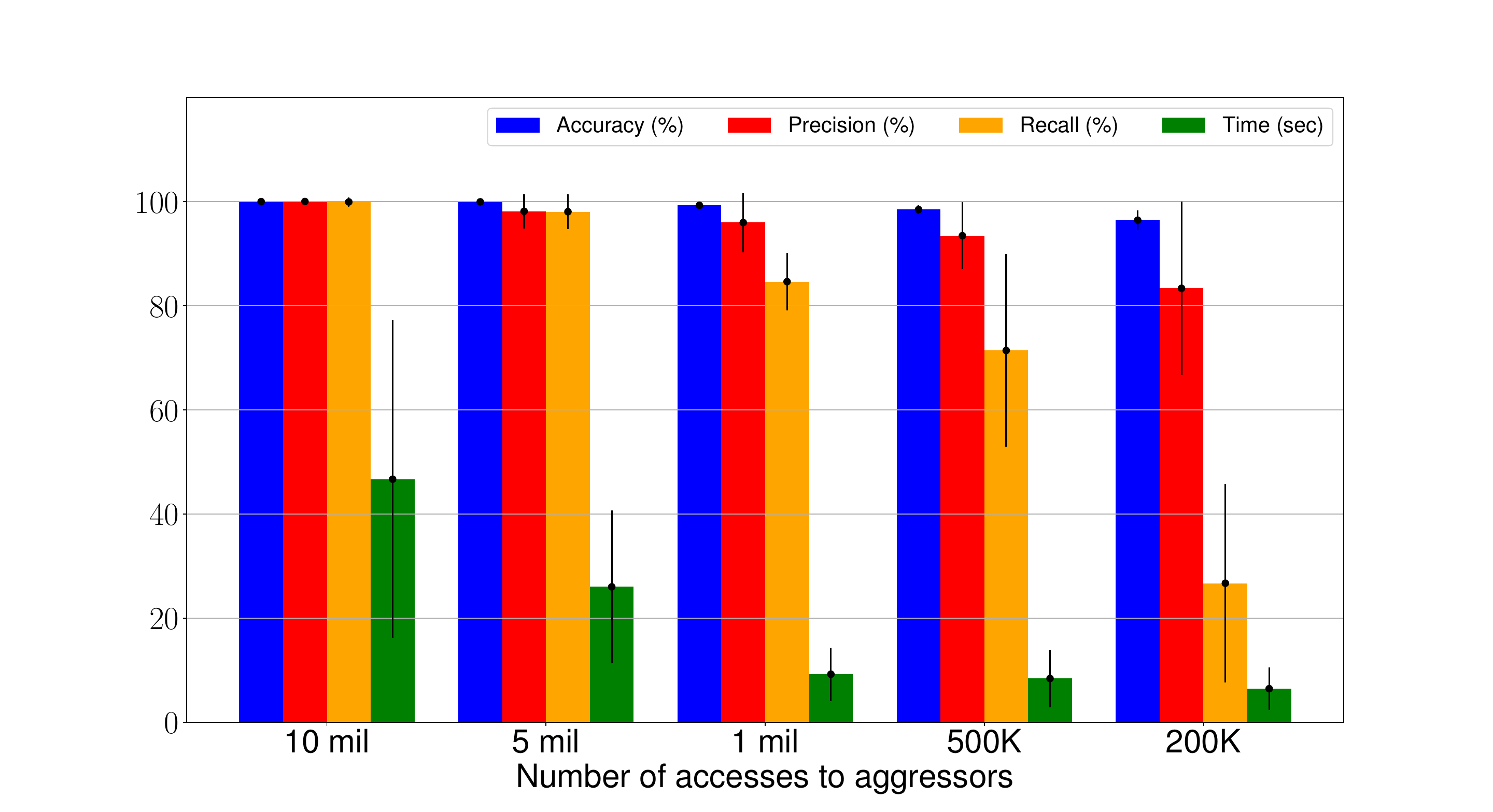}
                }
            }
            \vspace{-3mm}
            \caption{Plots showing the variation in accuracy, precision, recall and time
            elapsed to extract fingerprints when varying the number of accesses to trigger bit
            flips and varying the number of times we repeat the hammering sweep operation.\\}
            \label{fig:eval_eff}
        \end{center}
    \end{figure*}
We present the highest accuracy attained, with the corresponding precision and 
recall for all three experiments in Table \ref{fig:eval_stability}. The table shows
a drastically lower value of 50\% recall for the experiment where we 
re-seated the DIMM. The other two experiments show perfect precision and
recall. These observations support our hypothesis that the instability 
in \system's fingerprints is a result of re-seating the DIMMs. 

We suspect that minute differences in alignment when 
re-seating DIMMs change the electrical characteristics (resistance, capacitance,
inductance, etc) of the wires in the DRAM chips, which ultimately changes the set
of bits that flip.
Consistent with prior research \cite{rowhammer-reproduction}, these results 
show that the set of bits that flip (and correspondingly the fingerprints extracted
by \system) are a property of the entire system being subjected to Rowhammer 
and not just the DRAM devices. 
%Thus, we 
%can expect the fingerprints extracted by \system to be stable in the wild 
%since users rarely re-seat their DIMMs.
\vspace{-3mm}
\subsection{How long does \system take to extract fingerprints?}
\label{sec:eval_eff}
Efficiency quantifies the time taken to extract
fingerprints \cite{clock-skew}. When accessing aggressors 10,000,000 times to execute
Rowhammer, the hammering sweep operation takes an average of 20 seconds on a
single 2 MB chunk, which further increases when repeating the operation to account 
for non-determinism. Even in the limiting case where we have enough references 
to cover all 2 MB chunks in a given DIMM, \system takes almost 3 minutes 
to extract a fingerprint when accessing aggressors 10,000,000 times and 
repeating the hammering sweep 8 times. Such a long duration for fingerprinting
would only be acceptable in certain use cases where a user keeps an
application open for a long time, such as a gaming application or a video
streaming application. We can improve \system's efficiency by reducing the
number of times we access aggressors to trigger bit flips or reducing the
number of times we repeat the hammering sweep operation. Employing either 
approach is subject to making sure that they do not significantly degrade 
\system's fingerprinting accuracy.

In this section, we present a comprehensive analysis of 
the trade-off between fingerprint accuracy and efficiency
when running \system. Concretely, we present the fingerprint
accuracy and the average time taken to extract a fingerprint 
from one 2 MB chunk across 15  different configurations on 
all the DIMMs in our test bed. These 15 configurations differ
in terms of the number of times we access the aggressors when
executing Rowhammer and the number of times we repeat the
hammering sweep operation. We consider 5 different
values for the number of accesses to each aggressor ranging from
10 million accesses to 200,000 accesses and 3 different values
for the number of times we repeat the hammering sweep operation
(8 times, 4 times and 2 times).

Figure \ref{fig:eval_eff} shows the accuracy, precision, recall and 
time elapsed in all 15 configurations. These results
show that we  can drastically reduce the time taken
to extract fingerprints while maintaining high
accuracy. For example, accessing aggressors 1,000,000 times and repeating the hammering
sweep operation twice only takes 9.92 seconds, 
while still maintaining 95.96\% precision and 84.61\% recall
(99.27\% accuracy). In this configuration, \system is suitable as a 
second factor of authentication \cite{2fa-time, daredevil}.

To further decrease the time taken to extract
fingerprints, we ran an experiment where we hammered half the rows in each 2 MB chunk instead of all rows in the hammering sweep operation. With this optimization, it took us
4.54 seconds to extract fingerprints corresponding to 95.33\% precision and 83.84\% recall (99.09\% accuracy). %Fingerprinters can further reduce the time taken to extract fingerprints by further confining the hammering sweep to fewer rows as long as they can maintain high accuracy.
Fingerprinters can further constrain the hammering sweep to fewer rows or 
employ multi-back hammering \cite{sledgehammer} to further decrease the time 
taken to extract
fingerprints.
%Multi-bank hammering proposed by SledgeHammer \cite{sledgehammer} can also be used to further decrease the time taken to
%extract fingerprints.
\vspace{-3mm}
\subsection{Can \system extract robust fingerprints in presence of external factors?}
\label{sec:eval_freq}
We evaluate the robustness of \system's fingerprints to 
external factors (outside the control of the fingerprinter) that can influence 
the behavior of bit flips. Concretely,
we evaluate \system's robustness in context
of CPU frequency since fewer bits flip at lower
frequencies. CPU frequencies are subject to change since some
CPU governors (e.g., ondemand \cite{governors})
dynamically scale the CPU frequency based on the
CPU load. This load depends on other applications running 
on a user's device and cannot be controlled by the 
fingerprinter. Matching fingerprints extracted at
different frequencies 
corresponds to matching 
fingerprints when running
other applications with
different loads.

We first extract fingerprints (bit flip distributions) on all DIMMs in 
our test bed when running the CPU at its highest frequency of 3600 MHz. 
Then, we extract fingerprints from the same DIMMs when running the CPU 
at a lower frequency of 2800 MHz. Matching fingerprints across these
frequencies simulates an extreme scenario where we trigger bit flips 
in presence of different applications that exert different loads 
on the CPU. On average, we report 2 orders of magnitude difference in 
the number of bit flips that trigger at the two frequencies. When 
matching fingerprints across these two frequencies, \system achieves 
an accuracy of 99.09\%, corresponding to a precision of 94.2\% and 
recall of 81.56\%.  Thus, \system only suffers a modest drop in 
fingerprint accuracy when matching fingerprints extracted at different 
frequencies. 
 
Importantly, these results also demonstrate the impact of employing
hammering patterns that produce the most number of bit flips. We repeated
the same experiment by choosing a hammering pattern that triggered fewer
bit flips in the templating phase. We notice that even when running at the 
highest frequency of 3600 MHz, this pattern triggers bit flips on 
only 10 DIMMs. On these DIMMs, this pattern did not trigger any
bit flips at the lower frequency of 2800 MHz, resulting in 9\% recall when 
comparing fingerprints. %On the same set of 
%DIMMs, the pattern that triggered the most number of bit flips in the 
%templating phase has a fingerprint accuracy of 95.92\%, precision of 100\% 
%and recall of 71.43\%.

We also compared fingerprints while running certain common applications 
in the background. Concretely, on a subset of 10 random identical DIMMs, 
we extracted fingerprints when running 1) no background applications, 2) 
running LibreOffice Writer, and 3) running YouTube on Mozilla Firefox in 
the background. We report 100\% accuracy when matching these fingerprints 
against each other. We did not evaluate the robustness of fingerprints to
changes in temperature \cite{rowhammer-puf} since the ambient temperature
did not change significantly over the course of our experiments.

\vspace{-5mm}
\subsection{How does \system compare against prior research?}
\label{app:eval_baseline}
In this section, we compare \system's approach of hammering the same
chunk multiple times and comparing probability distributions to match fingerprints 
against the approach of hammering once and comparing the Jaccard similarity of 
the set of bit flips as proposed by prior research \cite{rowhammer-puf, fp-hammer}. 
When accessing aggressors 1 million times and repeating the hammering sweep operation 4 times, \system has an accuracy of 99.41\%, corresponding to 96.33\% precision and 89.93\%
recall. The same number of accesses to aggressors on the same set of DIMMs, leads to Jaccard similarity 98.01\% accuracy, corresponding to 89.59\% precision 64.64\% recall.

Furthermore, in presence of external factors, such as the CPU frequency, 
\system reports 99.09\% accuracy, 94.2\% precision and 81.56\% recall. 
On the same DIMMs, Jaccard index yields 96.3\% accuracy, 100\% precision and 
a significantly low 20.09\% recal. These results show that \system's approach 
of hammering chunks multiple times allows fingerprinters to account for the non-deterministic behavior of bit flips.
\section{Discussion}
\label{sec:disc}
\begin{comment}

\end{comment}
\subsection{Potential defenses against \system}
\textbf{Eliminating Rowhammer}
To the best of our knowledge, there exists no defense that can completely overcome Rowhammer. However, if such a defense were to be developed, it would also overcome \system since \system relies on observing bit flips produced by Rowhammer for fingerprinting.

\textbf{Restricting access to contiguous rows within a bank}
Any memory configuration that prevents access to contiguous rows 
within a single bank of a DIMM can be used to defend against \system.
For example, changing the DRAM address mapping such that the row bits
are not present within the lower order bits would ensure that fingerprinters
cannot pick appropriate aggressor rows to trigger Rowhammer without being 
able to allocate large amounts of contiguous memory. 
Such a mapping would practically make it impossible to execute 
double-sided Rowhammer, which is more reliable than single-sided
Rowhammer in producing bit flips. More importantly, even if a 
fingerprinter is able to trigger bit flips using single-sided
Rowhammer, they cannot observe them since bit flips are typically
triggered in rows that are adjacent to those being triggered.
However, it is difficult to change the mapping in the memory
controller of legacy devices which currently have their own
DRAM mapping. Changing the mapping in software would incur
significant performance overhead. Even with the proposed mapping,
fingerprinters may still be able to extract fingerprints if they
get fortuitous with memory allocation that grants them access to
contiguous rows in a bank.

%\textbf{Inapplicability of common defenses} We discuss how
%standard fingerprinting defenses (e.g., normalization, %obfuscation etc)
%and standard mitigations against Rowhammer (e.g., TRR, ECC etc) %are
%insufficient to overcome \system in Appendix %\ref{app:inapplicability}.

\subsection{Extension to web and mobile}
\label{sec:web-extension}
Executing Rowhammer, which is a prerequisite to use \system requires
access to contiguous memory and fast, uncached memory access ($\S$\ref{sec:rh}).
In this section, we discuss ways to satisfy these requirements on web and mobile 
systems to execute Rowhammer. Upon executing Rowhammer, fingerprinters can 
match bit flip distributions ($\S$\ref{sec:design}) on the 
resultant bit flips for fingerprinting.

Both Rowhammer.js \cite{rowhammer-js} and SMASH \cite{smash-rowhammer} describe 
ways to allocate contiguous memory from JavaScript using Transparent Huge Pages.
SMASH also describes self-evicting patterns that extend the many-sided hammering 
patterns proposed by TRRespass \cite{trrespass} to overcome TRR and trigger 
Rowhammer from the browser. However, TRResspass was unable to trigger bit 
flips on the DIMMs in our test bed, even after running their fuzzer for two 
weeks. We can also use Blacksmith's fuzzer \cite{blacksmith-code} to discover 
such patterns, although porting Blacksmith to JavaScript is non-trivial because
we cannot guarantee precisely accessing aggressors at particular points of
time within a refresh interval without explicitly flushing the cache.

Most recently, Sledgehammer \cite{sledgehammer} proposed a novel method to
allocate contiguous memory without relying on Transparent Huge Pages to trigger
bit flips from the browser. Drammer \cite{drammer} describes ways to allocate contiguous memory on
mobile platforms such as Android. Techniques proposed by SMASH and SledgeHammer 
to flush the cache can potentially be applied on mobile too to ensure fast, 
uncached memory access to execute Rowhammer.

The main contribution of our work, \system, is exploiting the distribution
of bit flips for fingerprinting. In this paper, we trigger Rowhammer bit flips 
using native code on desktop systems, but with considerable engineering effort, 
the contributions of \system can be extended to web and mobile systems by 
leveraging techniques proposed by Drammer, SMASH, Rowhammer.js and SledgeHammer.

\begin{figure*}[t]
\centering
   \includegraphics[width=2\columnwidth]{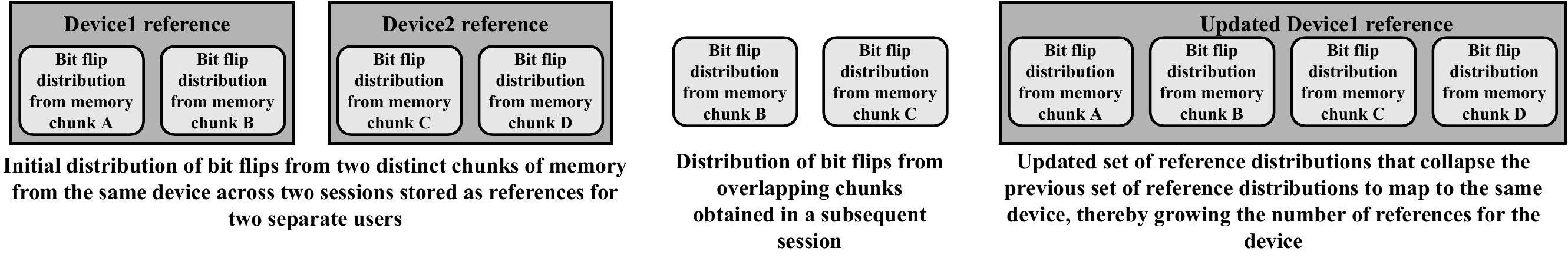}
   \vspace{-3mm}
  \caption{Incrementally building up references.
  In this figure, we assume that the fingerprinters first obtained bit flip distributions
  of two separate sets of contiguous chunks of memory from the same device across two separate sessions. Each set contains the distribution of bit flips observed on two such chunks. Since the distribution of bit flips on each chunk is unique, the fingerprinters treat these sets as belonging to two different devices. When they subsequently obtain bit flip distributions from chunks that overlap across both sets, they collapse them into a single reference that pertains to the same device. \\}
  \label{fig:incremental_references}
\end{figure*}
\subsection{Applications of \system}
In this section, we discuss applications of \system, focusing on potential
pitfalls and ways to overcome them. The unique and stable fingerprints 
extracted by \system can be used for authentication. Even \system's
least efficient configuration that takes almost 3 minutes to extract
fingerprints can be used for cheat detection in multiplayer gaming 
since users tend to keep gaming applications open for extended periods
of time.  Shorter variants of \system that take 4.5 seconds to extract
fingerprints are comparable to secondary forms of authentication such 
as CAPTCHAs or SMS-based authentication. \system provides better user
experience over these factors since \system can run seamlessly in the
background without requiring user interaction. Even shorter variants of
\system can potentially be used as the primary form of authentication.

However, using \system for authentication comes with non-zero risk
to benign users. \system could potentially crash another application or the OS
by flipping a sensitive bit reserved for a different application or
reserved by the OS. While background applications did not crash in 
our experiments ($\S$\ref{sec:eval_freq}), we cannot rule out their possibility. 
Although they were rare, we also report instances of devices crashing in 
our experiments, potentially due to bit flips in memory reserved for the OS.

Hammering those rows that are not at the edge of 2 MB chunks 
potentially addresses this concern. Most bit flips are observed in the 
rows that are adjacent to the aggressor rows and the number of bit 
flips successively decrease on rows that are farther away 
\cite{half-double}. Thus, this solution would reduce the chance of 
unintended bit flips by ensuring that most bit flips are confined 
to the desired 2 MB block. However,  this solution does not completely 
eliminate the possibility of unintended bit flips. OS vendors can help 
mitigate these concerns by allocating specific DRAM rows for \system 
that are far away from the rows allocated to the OS or other applications.

Another risk presented by \system is that it could wear out
memory modules if it is used to constantly trigger bit flips
for fingerprinting. FP-Rowhammer’s approach of triggering bit flips
with fewer accesses to aggressors helps mitigate this concern.
Such concerns can also be mitigated by only employing other
fingerprinting techniques for the common cases and sparingly
employing \system to only handle the critical cases.

\subsection{Incrementally building up references}
We discuss an alternate design of \system that
does not leave a distinct footprint by having to allocate multiple 
2 MB chunks in the initial stages before reaching 
the limiting case ($\S$\ref{sec:matching_phase}).

Fingerprinters can confine their hammering to fewer chunks (say, 2 chunks) per session to 
incrementally build up their references. In this case, fingerprinters would initially have 
multiple sets of references for devices without being able to link references that come 
from the same device. With this approach, fingerprinters would also not be able to fingerprint 
devices during their initial sessions. Eventually, after aggregating distributions from 
devices across multiple sessions, fingerprinters would be able to link sets of references to 
the same device and thereon fingerprint devices across all sessions. We demonstrate this approach with an example in Figure~\ref{fig:incremental_references}. 
Say, during a user's first session when running \system on a new device, the fingerprinter obtains the 
distribution of bit flips in 2 distinct chunks of 2 MB of memory, chunk A and chunk 
B. Since each chunk has a unique distribution of bit flips, neither chunk would match 
any existing reference chunk known to the fingerprinter. The fingerprinter would create a fresh 
reference for these two chunks. In the next session on that device, say the fingerprinter 
obtains the distributions of 2 other chunks, chunk C and chunk D. Again, since the 
distribution of bit flips in each chunk is unique, the fingerprinter will not be able to 
identify that this session corresponds to the same device, and would store them as a 
separate reference. In the next session with the user, say the fingerprinter obtains the 
bit flip distribution to chunk B and chunk C. Now, the fingerprinter would be able to 
fingerprint the device, and also combine the references containing chunks A and B 
with the references containing chunks C and D as chunks obtained from the same device. 
In a subsequent session on the same device, say the fingerprinter obtains the bit flip 
distributions to chunk C and chunk E. For this session too, the fingerprinters would be able to 
fingerprint the device and also extend their references for the user to contain the 
distribution of chunk E. Thus, when incrementally building up references, fingerprinters 
would be able to fingerprint users without leaving behind a distinct memory footprint. However, fingerprinters would have to give up being able to fingerprint devices during their initial sessions.
\section{Related Work}
\label{sec:related_work}
Drammer \cite{drammer} proposes a technique to overcome 
the operating system's abstractions to force a victim to allocate 
memory in a region that is susceptible to Rowhammer. However, 
to trigger the same bit flip again, their proposed technique requires 
the overall memory layout to remain unchanged (no allocation/deallocation 
of memory by other processes), which makes it impractical for fingerprinting.
Researchers have also leveraged memory deduplication and MMU paravirtualization 
to overcome memory restrictions and trigger bit flips on memory allocated to 
a victim VM on cloud machines \cite{ffs, bit_flip_cloud_flop}. However, 
these capabilities are either not enabled or unavailable on most end-user devices.

TRRespass \cite{trrespass} and Blacksmith \cite{blacksmith} 
introduce fuzzers that discover hammering patterns to overcome 
TRR and trigger bit flips. However, fingerprinters get limited insights on employing these 
patterns to trigger bit flips on a large number of DIMMs since 
both papers do not provide guidance on which patterns 
to employ beyond those that produce bit flips.

\begin{comment}
Researchers have also looked at other hardware components to extract device fingerprints. CryptoFP \cite{clock-skew} exploits differences in clock cycles emerging from small variations in the quartz crystals in oscillators to present a host-based (when the fingerprinter can run native code) and a web-based (when the fingerprinter can only run JavaScript on a browser) fingerprinting technique. DeMiCPU \cite{demicpu} proposes
the use of a specialized sensor to capture differences in
electromagnetic radiation to extract device fingerprints for authentication. SENSORID \cite{sensor-id} captures calibration errors in motion sensors on mobile phones for fingerprinting. DrawnApart \cite{drawn_apart-gpu} exploits differences in the time taken by different GPUs for rendering to link software fingerprint evolutions.
\end{comment}
\vspace{-1mm}
\section{Conclusion}
\label{sec:conclusion}
We presented \system to extract unique and stable fingerprints even for devices with identical hardware and software configurations. 
To this end, \system leverages Rowhammer to capture the side-effects of process variation in the underlying manufacturing process of memory modules. 
\system's design involves a novel sampling strategy to overcome memory allocation constraints, identification of effective hammering patterns to bypass Rowhammer mitigations and trigger bit flips at scale, and handling non-deterministic bit flips through multiple hammering iterations and divergence analysis of probability distributions.
Our evaluation of \system on 98 DIMMs across 6 sets of identical DRAM modules from two manufacturers showed that it can extract high entropy and stable fingerprints with an overall accuracy of 99.91\% while being robust and efficient.  
\system cannot be trivially mitigated without fundamentally fixing the underlying the Rowhammer vulnerability, which -- despite existing countermeasures -- is expected to escalate as the density of DRAM chips increases in the future.

\bibliographystyle{plain}
\bibliography{paper.bib}

\begin{thebibliography}{10}

\bibitem{abramson1970more}
Morton Abramson and WOJ Moser.
\newblock More birthday surprises.
\newblock {\em The American Mathematical Monthly}, 77(8):856--858, 1970.

\bibitem{norm-entropy}
Nampoina Andriamilanto, Tristan Allard, Ga\"{e}tan Le~Guelvouit, and Alexandre Garel.
\newblock A large-scale empirical analysis of browser fingerprints properties for web authentication.
\newblock {\em ACM Trans. Web}, 16(1), sep 2021.

\bibitem{idfa}
{Apple}.
\newblock {User Privacy and Data Use}.
\newblock \url{https://developer.apple.com/app-store/user-privacy-and-data-use/}.

\bibitem{anvil}
Zelalem~Birhanu Aweke, Salessawi~Ferede Yitbarek, Rui Qiao, Reetuparna Das, Matthew Hicks, Yossi Oren, and Todd Austin.
\newblock Anvil: Software-based protection against next-generation rowhammer attacks.
\newblock In {\em Proceedings of the Twenty-First International Conference on Architectural Support for Programming Languages and Operating Systems}, ASPLOS '16, page 743–755, New York, NY, USA, 2016. Association for Computing Machinery.

\bibitem{intel-patent}
Kuljit~S. Bains and John~B. Halbert.
\newblock {Distributed row hammer tracking}, 2012.
\newblock {US Patent US20140095780A1}.

\bibitem{iosfingerprinting}
{Benjamin Seufert}.
\newblock {Apple to Ad Tech: ``Fingerprinting is Never Allowed''}.
\newblock \url{https://mobiledevmemo.com/apple-to-adtech-fingerprinting-is-never-allowed/}.

\bibitem{dedup-est-machina}
Erik Bosman, Kaveh Razavi, Herbert Bos, and Cristiano Giuffrida.
\newblock Dedup est machina: Memory deduplication as an advanced exploitation vector.
\newblock In {\em 2016 IEEE Symposium on Security and Privacy (SP)}, pages 987--1004, 2016.

\bibitem{windows-rowhammer}
Erik Bosman, Kaveh Razavi, Herbert Bos, and Cristiano Giuffrida.
\newblock Dedup est machina: Memory deduplication as an advanced exploitation vector.
\newblock In {\em 2016 IEEE Symposium on Security and Privacy (SP)}, pages 987--1004, 2016.

\bibitem{brave-canvas}
{Brave}.
\newblock {Fingerprinting Protection Mode}.
\newblock \url{https://github.com/brave/browser-laptop/wiki/Fingerprinting-Protection-Mode}.

\bibitem{farbling}
{Brave Privacy Team}.
\newblock {Fingerprinting defenses 2.0}.
\newblock \url{https://brave.com/privacy-updates/4-fingerprinting-defenses-2.0/}.

\bibitem{demicpu}
Yushi Cheng, Xiaoyu Ji, Juchuan Zhang, Wenyuan Xu, and Yi-Chao Chen.
\newblock Demicpu: Device fingerprinting with magnetic signals radiated by cpu.
\newblock In {\em Proceedings of the 2019 ACM SIGSAC Conference on Computer and Communications Security}, CCS '19, page 1149–1170, New York, NY, USA, 2019. Association for Computing Machinery.

\bibitem{rowhammer-ecc}
Lucian Cojocar, Kaveh Razavi, Cristiano Giuffrida, and Herbert Bos.
\newblock Exploiting correcting codes: On the effectiveness of ecc memory against rowhammer attacks.
\newblock In {\em 2019 IEEE Symposium on Security and Privacy (SP)}, pages 55--71, 2019.

\bibitem{smash-rowhammer}
Finn de~Ridder, Pietro Frigo, Emanuele Vannacci, Herbert Bos, Cristiano Giuffrida, and Kaveh Razavi.
\newblock {SMASH}: Synchronized many-sided rowhammer attacks from {JavaScript}.
\newblock In {\em 30th USENIX Security Symposium (USENIX Security 21)}, pages 1001--1018. USENIX Association, August 2021.

\bibitem{daredevil}
Zainul~Abi Din, Hari Venugopalan, Henry Lin, Adam Wushensky, Steven Liu, and Samuel~T. King.
\newblock Doing good by fighting fraud: Ethical anti-fraud systems for mobile payments, 2021.

\bibitem{governors}
{Dominik Brodowski, Nico Golde, Rafael J. Wysocki and Viresh Kumar}.
\newblock {Linux CPUFreq CPUFreq Governor}.
\newblock \url{https://www.kernel.org/doc/Documentation/cpu-freq/governors.txt}.

\bibitem{easy-anti-cheat}
{easy ANTI-CHEAT}.
\newblock {Don't bear with the cheaters}.
\newblock \url{https://www.easy.ac/en-US}.

\bibitem{browser-uniqueness}
Peter Eckersley.
\newblock How unique is your web browser?
\newblock In {\em Proceedings of the 10th International Conference on Privacy Enhancing Technologies}, PETS'10, page 1–18, Berlin, Heidelberg, 2010. Springer-Verlag.

\bibitem{pandaboard}
{elinux}.
\newblock {PandaBoard}.
\newblock \url{https://elinux.org/PandaBoard}.

\bibitem{fingerprintjs-github}
{FingerprintJS}.
\newblock {FingerprintJS}.
\newblock \url{https://github.com/fingerprintjs/fingerprintjs}.

\bibitem{fpjs-android}
{FingerprintJS}.
\newblock {FingerprintJS Android}.
\newblock \url{https://github.com/fingerprintjs/fingerprintjs-android}.

\bibitem{fpjs-ios}
{FingerprintJS}.
\newblock {FingerprintJS iOS}.
\newblock \url{https://github.com/fingerprintjs/fingerprintjs-ios}.

\bibitem{trrespass}
Pietro Frigo, Emanuele Vannacc, Hasan Hassan, Victor~van der Veen, Onur Mutlu, Cristiano Giuffrida, Herbert Bos, and Kaveh Razavi.
\newblock Trrespass: Exploiting the many sides of target row refresh.
\newblock In {\em 2020 IEEE Symposium on Security and Privacy (SP)}, pages 747--762, 2020.

\bibitem{rowhammer-reproduction}
Lukas Gerlach, Fabian Thomas, Robert Pietsch, and Michael Schwarz.
\newblock A rowhammer reproduction study using the blacksmith fuzzer.
\newblock In Gene Tsudik, Mauro Conti, Kaitai Liang, and Georgios Smaragdakis, editors, {\em Computer Security -- ESORICS 2023}, pages 62--79, Cham, 2024. Springer Nature Switzerland.

\bibitem{adid}
{Google}.
\newblock {Advertising ID}.
\newblock \url{https://support.google.com/googleplay/android-developer/answer/6048248}.

\bibitem{hlisa}
Daniel Go\ss{}en, Hugo Jonker, Stefan Karsch, Benjamin Krumnow, and David Roefs.
\newblock Hlisa: towards a more reliable measurement tool.
\newblock In {\em Proceedings of the 21st ACM Internet Measurement Conference}, IMC '21, page 380–389, New York, NY, USA, 2021. Association for Computing Machinery.

\bibitem{rowhammer-js}
Daniel Gruss, Cl\'{e}mentine Maurice, and Stefan Mangard.
\newblock Rowhammer.js: A remote software-induced fault attack in javascript.
\newblock In {\em Proceedings of the 13th International Conference on Detection of Intrusions and Malware, and Vulnerability Assessment - Volume 9721}, DIMVA 2016, page 300–321, Berlin, Heidelberg, 2016. Springer-Verlag.

\bibitem{utrr}
Hasan Hassan, Yahya~Can Tugrul, Jeremie~S. Kim, Victor van~der Veen, Kaveh Razavi, and Onur Mutlu.
\newblock Uncovering in-dram rowhammer protection mechanisms:a new methodology, custom rowhammer patterns, and implications.
\newblock In {\em MICRO-54: 54th Annual IEEE/ACM International Symposium on Microarchitecture}, MICRO '21, page 1198–1213, New York, NY, USA, 2021. Association for Computing Machinery.

\bibitem{gddr5}
{Hynix Semiconductor}.
\newblock {Datasheet for 1Gb (32Mx32) GDDR5 SGRAM H5GQ1H24AFR}.
\newblock Technical Report H5GQ1H24AFR, 2009.

\bibitem{iqbal2021fingerprinting}
Umar Iqbal, Steven Englehardt, and Zubair Shafiq.
\newblock Fingerprinting the fingerprinters: Learning to detect browser fingerprinting behaviors.
\newblock In {\em 2021 IEEE Symposium on Security and Privacy (SP)}, pages 1143--1161. IEEE, 2021.

\bibitem{memorybook}
Bruce Jacob, Spencer Ng, and David Wang.
\newblock {\em Memory Systems: Cache, DRAM, Disk}.
\newblock Morgan Kaufmann Publishers Inc., San Francisco, CA, USA, 2007.

\bibitem{blacksmith}
Patrick Jattke, Victor Van Der~Veen, Pietro Frigo, Stijn Gunter, and Kaveh Razavi.
\newblock Blacksmith: Scalable rowhammering in the frequency domain.
\newblock In {\em 2022 IEEE Symposium on Security and Privacy (SP)}, pages 716--734, 2022.

\bibitem{blacksmith-code}
Patrick Jattke, Victor van~der Veen, Pietro Frigo, Stijn Gunter, and Kaveh Razavi.
\newblock {{BLACKSMITH}}: Rowhammering in the {{Frequency Domain}}, 2022-05.
\newblock \url{https://github.com/comsec-group/blacksmith}.

\bibitem{zenhammer}
Patrick Jattke, Max Wipfli, Flavien Solt, Michele Marazzi, Matej Bölcskei, and Kaveh Razavi.
\newblock Zenhammer: Rowhammer attacks on amd zen-based platforms, 2024.

\bibitem{ddr5}
{JEDEC}.
\newblock {DDR5 SDRAM}.
\newblock Technical Report {JESD79-5B}, {August} 2022.

\bibitem{sledgehammer}
Ingab Kang, Walter Wang, Jason Kim, Stephan van Schaik, Youssef Tobah, Daniel Genkin, Andrew Kwong, and Yuval Yarom.
\newblock Sledgehammer: Amplifying rowhammer via bank-level parallelism, 2024.

\bibitem{thp}
{kernel development community}.
\newblock {Transparent Hugepage Support}.
\newblock \url{https://www.kernel.org/doc/html/latest/admin-guide/mm/transhuge.html}.

\bibitem{pagemap}
{kernel.org}.
\newblock {pagemap, from the userspace perspective}.
\newblock \url{https://www.kernel.org/doc/Documentation/vm/pagemap.txt}.

\bibitem{kim-rowhammer}
Yoongu Kim, Ross Daly, Jeremie Kim, Chris Fallin, Ji~Hye Lee, Donghyuk Lee, Chris Wilkerson, Konrad Lai, and Onur Mutlu.
\newblock Flipping bits in memory without accessing them: An experimental study of dram disturbance errors.
\newblock {\em SIGARCH Comput. Archit. News}, 42(3):361–372, jun 2014.

\bibitem{half-double}
Andreas Kogler, Jonas Juffinger, Salman Qazi, Yoongu Kim, Moritz Lipp, Nicolas Boichat, Eric Shiu, Mattias Nissler, and Daniel Gruss.
\newblock {Half-Double}: Hammering from the next row over.
\newblock In {\em 31st USENIX Security Symposium (USENIX Security 22)}, pages 3807--3824, Boston, MA, August 2022. USENIX Association.

\bibitem{drawn_apart-gpu}
Tomer Laor, Naif Mehanna, Antonin Durey, Vitaly Dyadyuk, Pierre Laperdrix, Cl{\'{e} }mentine Maurice, Yossi Oren, Romain Rouvoy, Walter Rudametkin, and Yuval Yarom.
\newblock {DRAWN} {APART} : A device identification technique based on remote {GPU} fingerprinting.
\newblock In {\em Proceedings 2022 Network and Distributed System Security Symposium}. Internet Society, 2022.

\bibitem{laperdrix2020browser}
Pierre Laperdrix, Nataliia Bielova, Benoit Baudry, and Gildas Avoine.
\newblock Browser fingerprinting: A survey.
\newblock {\em ACM Transactions on the Web (TWEB)}, 14(2):1--33, 2020.

\bibitem{fp-hammer}
Dawei Li, Di~Liu, Yangkun Ren, Ziyi Wang, Yu~Sun, Zhenyu Guan, Qianhong Wu, and Jianwei Liu.
\newblock Fphammer: A device identification framework based on dram fingerprinting, 2023.

\bibitem{dramsim3}
Shang Li, Zhiyuan Yang, Dhiraj Reddy, Ankur Srivastava, and Bruce Jacob.
\newblock Dramsim3: A cycle-accurate, thermal-capable dram simulator.
\newblock {\em IEEE Computer Architecture Letters}, 19(2):106--109, 2020.

\bibitem{lspci}
{man7.org}.
\newblock {lspci(8) — Linux manual page}.
\newblock \url{https://man7.org/linux/man-pages/man8/lspci.8.html}.

\bibitem{micron-trr}
{Micron}.
\newblock {DDR4 SDRAM}.
\newblock \url{https://www.micron.com/-/media/client/global/documents/products/data-sheet/dram/ddr4/8gb_ddr4_sdram.pdf}.

\bibitem{microLPDDRx}
{Micron Technology}.
\newblock {TN-46-12: Mobile DRAM Power-Saving Features and Power Calculations}.
\newblock Technical Report {TN46\_12}, 2005.

\bibitem{ddr2}
{Micron Technology}.
\newblock {DDR2 SDRAM }.
\newblock Technical Report MT47H512M4,MT47H256M8,MT47H128M16, 2006.

\bibitem{ddr3}
{Micron Technology}.
\newblock {TN-41-01: Calculating Memory System Power for DDR3}.
\newblock Technical Report {TN41\_01DDR3}, January 2007.

\bibitem{ddr3l}
{Micron Technology}.
\newblock {1.35V DDR3L SDRAM SODIMM}.
\newblock Technical Report {MT16KTF51264HZ, MT16KTF1G64HZ}, 2011.

\bibitem{ddr4}
{Micron Technology,}.
\newblock {DDR4 SDRAM}.
\newblock Technical Report {MT40A2G4, MT40A1G8, MT40A512M16}, 2015.

\bibitem{gddr6}
{Micron Technology}.
\newblock {TN-ED-03: GDDR6: The Next-Generation Graphics DRAM }.
\newblock Technical Report TN-ED-03: GDDR6, 2017.

\bibitem{tor}
{Mike Perry, Erinn Clark, Steven Murdoch and Georg Koppen}.
\newblock {The Design and Implementation of the Tor Browser}.
\newblock \url{https://2019.www.torproject.org/projects/torbrowser/design/}.

\bibitem{cookie}
{Mozilla}.
\newblock {Using HTTP cookies}.
\newblock \url{https://developer.mozilla.org/en-US/docs/Web/HTTP/Cookies}.

\bibitem{canvas-defender}
{MultiLogin}.
\newblock {Hardware: Canvas}.
\newblock \url{https://docs.multilogin.com/l/en/article/7gNVYHcFKG-canvas}.

\bibitem{js-divergence}
{Notes on AI}.
\newblock Jensen--shannon divergence.
\newblock Accessed: 2024-09-12.

\bibitem{battery-fingerprint}
Lukasz Olejnik, Gunes Acar, Claude Castelluccia, and Claudia D{\'i}az.
\newblock The leaking battery - a privacy analysis of the html5 battery status api.
\newblock In {\em DPM/QASA@ESORICS}, 2015.

\bibitem{hennesyandpatterson}
David~A. Patterson and John~L. Hennessy.
\newblock {\em Computer Architecture: A Quantitative Approach}.
\newblock Morgan Kaufmann Publishers Inc., San Francisco, CA, USA, 1990.

\bibitem{drama}
Peter Pessl, Daniel Gruss, Cl{\'e}mentine Maurice, Michael Schwarz, and Stefan Mangard.
\newblock {DRAMA}: Exploiting {DRAM} addressing for {Cross-CPU} attacks.
\newblock In {\em 25th USENIX Security Symposium (USENIX Security 16)}, pages 565--581, Austin, TX, August 2016. USENIX Association.

\bibitem{stemming}
Gaston Pugliese, Christian Riess, Freya Gassmann, and Zinaida Benenson.
\newblock Long-term observation on browser fingerprinting: Users' trackability and perspective.
\newblock {\em Proceedings on Privacy Enhancing Technologies}, 2020:558--577, 05 2020.

\bibitem{ffs}
Kaveh Razavi, Ben Gras, Erik Bosman, Bart Preneel, Cristiano Giuffrida, and Herbert Bos.
\newblock Flip feng shui: Hammering a needle in the software stack.
\newblock In {\em 25th USENIX Security Symposium (USENIX Security 16)}, pages 1--18, Austin, TX, August 2016. USENIX Association.

\bibitem{reardon201950}
Joel Reardon, {\'A}lvaro Feal, Primal Wijesekera, Amit Elazari~Bar On, Narseo Vallina-Rodriguez, and Serge Egelman.
\newblock 50 ways to leak your data: An exploration of apps' circumvention of the android permissions system.
\newblock In {\em 28th USENIX Security Symposium (USENIX Security 19)}, pages 603--620, 2019.

\bibitem{dramsim2}
Paul Rosenfeld, Elliott Cooper-Balis, and Bruce Jacob.
\newblock Dramsim2: A cycle accurate memory system simulator.
\newblock {\em IEEE Computer Architecture Letters}, 10(1):16--19, 2011.

\bibitem{ddr4-samsung}
{Samsung Electronics}.
\newblock {DDR4 SDRAM}.
\newblock Technical report, 2014.

\bibitem{clock-skew}
Iskander Sanchez-Rola, Igor Santos, and Davide Balzarotti.
\newblock {Clock Around the Clock: Time-Based Device Fingerprinting}.
\newblock In {\em Proceedings of the 2018 ACM SIGSAC Conference on Computer and Communications Security}, CCS '18, page 1502–1514, New York, NY, USA, 2018. Association for Computing Machinery.

\bibitem{rowhammer-puf}
Andre Schaller, Wenjie Xiong, Nikolaos~Athanasios Anagnostopoulos, Muhammad~Umair Saleem, Sebastian Gabmeyer, Stefan Katzenbeisser, and Jakub Szefer.
\newblock Intrinsic rowhammer {PUFs}: Leveraging the rowhammer effect for improved security.
\newblock In {\em 2017 {IEEE} International Symposium on Hardware Oriented Security and Trust ({HOST})}. {IEEE}, may 2017.

\bibitem{decode_dimms}
{Sensirion}.
\newblock {EEPROM data decoder for SDRAM DIMM modules}.
\newblock \url{https://github.com/Sensirion/i2c-tools/blob/master/eeprom/decode-dimms}.

\bibitem{SHOCKLEY196135}
William Shockley.
\newblock Problems related to p-n junctions in silicon.
\newblock {\em Solid-State Electronics}, 2(1):35--67, 1961.

\bibitem{drammer}
Victor van~der Veen, Yanick Fratantonio, Martina Lindorfer, Daniel Gruss, Clementine Maurice, Giovanni Vigna, Herbert Bos, Kaveh Razavi, and Cristiano Giuffrida.
\newblock Drammer: Deterministic rowhammer attacks on mobile platforms.
\newblock In {\em Proceedings of the 2016 ACM SIGSAC Conference on Computer and Communications Security}, CCS '16, page 1675–1689, New York, NY, USA, 2016. Association for Computing Machinery.

\bibitem{fp-stalker}
Antoine Vastel, Pierre Laperdrix, Walter Rudametkin, and Romain Rouvoy.
\newblock Fp-stalker: Tracking browser fingerprint evolutions.
\newblock In {\em 2018 IEEE Symposium on Security and Privacy (SP)}, pages 728--741, 2018.

\bibitem{aragorn}
Hari Venugopalan, Zainul~Abi Din, Trevor Carpenter, Jason Lowe-Power, Samuel~T. King, and Zubair Shafiq.
\newblock Aragorn: A privacy-enhancing system for mobile cameras.
\newblock {\em Proc. ACM Interact. Mob. Wearable Ubiquitous Technol.}, 7(4), jan 2024.

\bibitem{2fa-time}
Catherine~S. Weir, Gary Douglas, Martin Carruthers, and Mervyn Jack.
\newblock User perceptions of security, convenience and usability for ebanking authentication tokens.
\newblock {\em Computers \& Security}, 28(1):47--62, 2009.

\bibitem{bit_flip_cloud_flop}
Yuan Xiao, Xiaokuan Zhang, Yinqian Zhang, and Radu Teodorescu.
\newblock One bit flips, one cloud flops: {Cross-VM} row hammer attacks and privilege escalation.
\newblock In {\em 25th USENIX Security Symposium (USENIX Security 16)}, pages 19--35, Austin, TX, August 2016. USENIX Association.

\end{thebibliography}

\begin{appendix}
    \section{Deriving the number of 2 MB chunks to access in each session to guarantee access to at 
    least one overlapping chunk across sessions}
    \label{app:math}
    Suppose a user's device has $N$ distinct 2 MB chunks of memory. As fingerprinters,
    we want to find $d$, the minimum number of distinct chunks to access in each session 
    to guarantee that at least one chunk overlaps between sessions. Once we have an overlapping
    chunk, we can use the unique distribution of bit flips on the chunk to identify the user's
    device.
    
    To solve for d, we set up equations that describe the probability that at least one 2 MB
    chunk would overlap across 2 sessions. Let $A$ be the event that at least one chunk overlaps
    between 2 sessions. For a given $N$, we want to find the minimum $d$ such that $P(A \mid N, d) \approx 1$

    $P(A \mid N, d) = 1 - P(\overline{A} \mid N, d)$

    Here, $P(\overline{A} \mid N, d)$ is the probability that no chunk overlapped in the two sessions.
    Calculating $P(\overline{A} \mid N, d)$ is easy since it is the probability of choosing $d$ chunks
    among $N$ chunks in one session multiplied by the probability of choosing $d$ chunks among $(N - d)$ chunks
    in the next session. Mathematically,

    $P(\overline{A} \mid N, d)$ = $N \choose d$/$N \choose d$ $\times$ $(N - d) \choose d$/$N \choose d$
    
    $\implies$$P(\overline{A} \mid N, d)$ = $(N - d) \choose d$/$N \choose d$
    
    $\implies$$P(\overline{A} \mid N, d)$ = $(N - d) \choose d$/$N \choose d$
    
    $\implies$$P(\overline{A} \mid N, d)$ = $\frac{(N-d)! (N-d)!}{N!(N-2d)!}$

    $\implies$$P(\overline{A} \mid N, d)$ = $\frac{(N-d)(N-d-1)\cdots(N-2d+1)}{N(N-1)\cdots(N-d+1)}$

    $\implies$$P(A \mid N, d)$ = 1 - $\frac{(N-d)(N-d-1)\cdots(N-2d+1)}{N(N-1)\cdots(N-d+1)}$

    Plugging in $N=512$ which represents $1$ GB of memory and $d=64$ yields a probability
    of $0.9998 \approx 1$. 

    Using the same formulation, we can define the following function
    that determines the number of chunks to sample to get an 
    overlapping chunk against the number of reference chunks:
    
    $P$ = 1 - ($N \choose S$/$(N - S) \choose d$ $\times$ $N \choose S$/$N \choose d$)

    where, $N$ refers to the total number of chunks, $S$ refers to the
    number of chunks previously sampled as part of the reference and $d$ refers to the number of chunks to be sampled in the current session to ensure that we have a probability of $P$ that at least one chunk overlaps with the reference.

\begin{figure}[!t]
    \centering
      \includegraphics[width=\columnwidth]{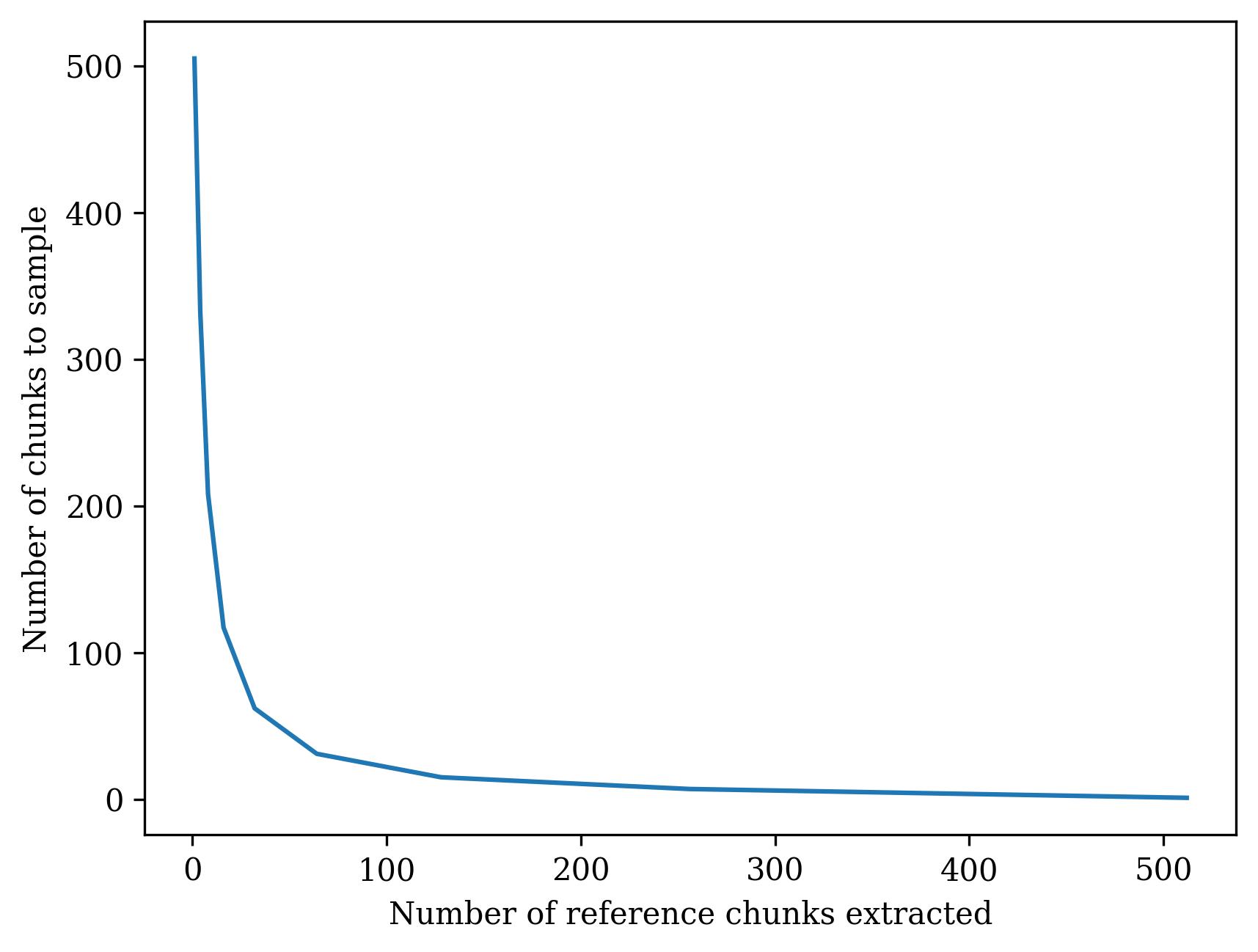}
      \vspace{-3mm}
      \caption{Plot shows the number of chunks to sample in one session to ensure at least one overlapping chunks against the number of reference chunks extracted. We see that the number of chunks to sample drops significantly as the number of reference chunks increases.}
      \label{fig:birthday}
    \end{figure}

    Once we setting $P=0.999$ and $N=512$ for $1$ GB of memory, we
    can calculate values of $d$ for varying values of $S$. We visualize this plot in Figure \ref{fig:birthday}.
    \vspace{-3mm}
    
    \section{Distirbution of JS divergence values across pairs of fingerprints on identical DIMMs from alternate manufacturer}
\begin{figure*}[tb]
        \begin{center}
            \mbox{ 
            \hspace{-1.0ex}
            \subfigure[Distribution of JS divergence values across all pairs of 
            fingerprints from 4 identical 2Rx8 DIMMs]
                {
                \label{fig:b_short_2rx8}
                \includegraphics[width=0.3\textwidth]{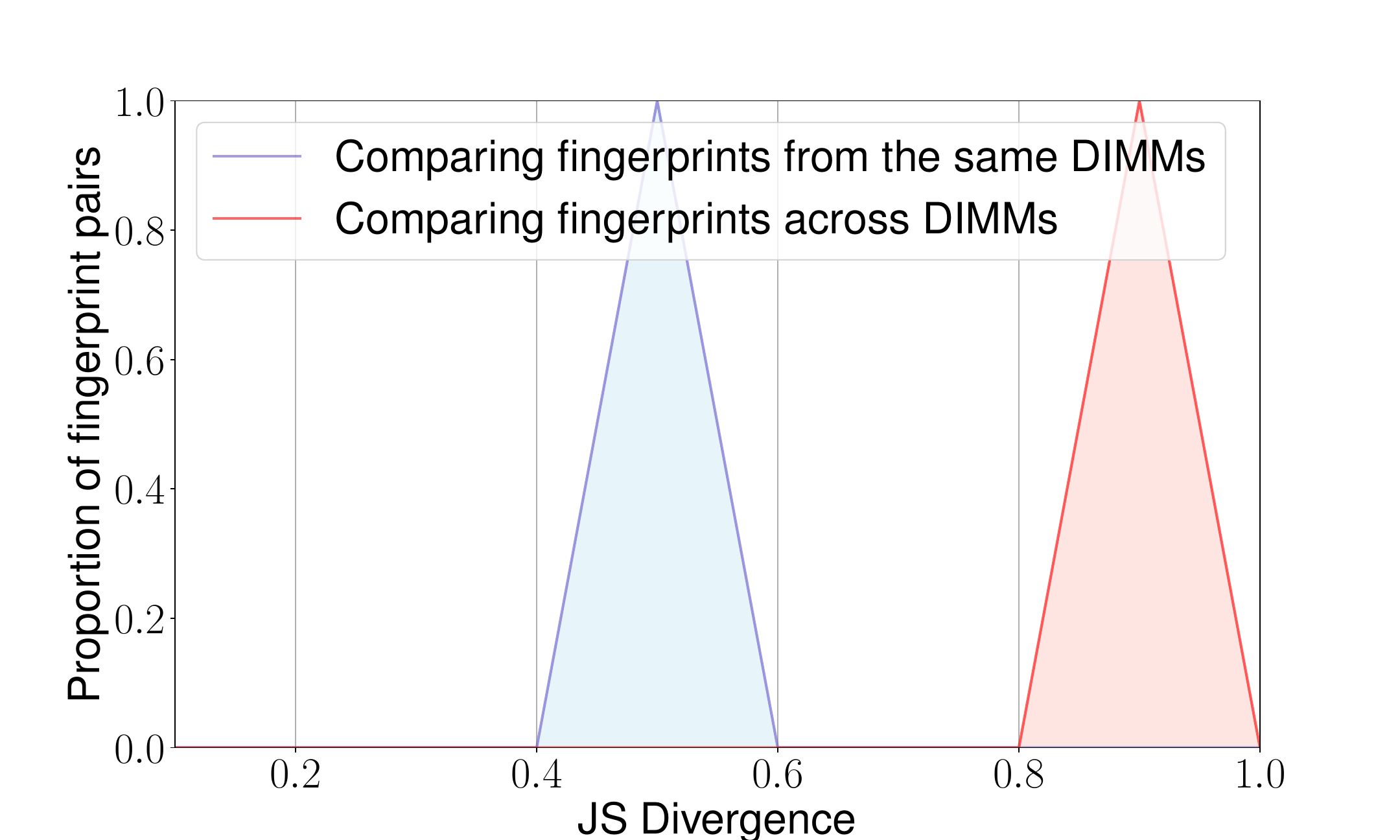}
                }
            \hspace{1.0ex}
            \subfigure[Distribution of JS divergence values across all pairs of 
            fingerprints from 10 identical 1Rx8 DIMMs]
                {
                \label{fig:b_short_1rx8}
                \includegraphics[width=0.3\textwidth]{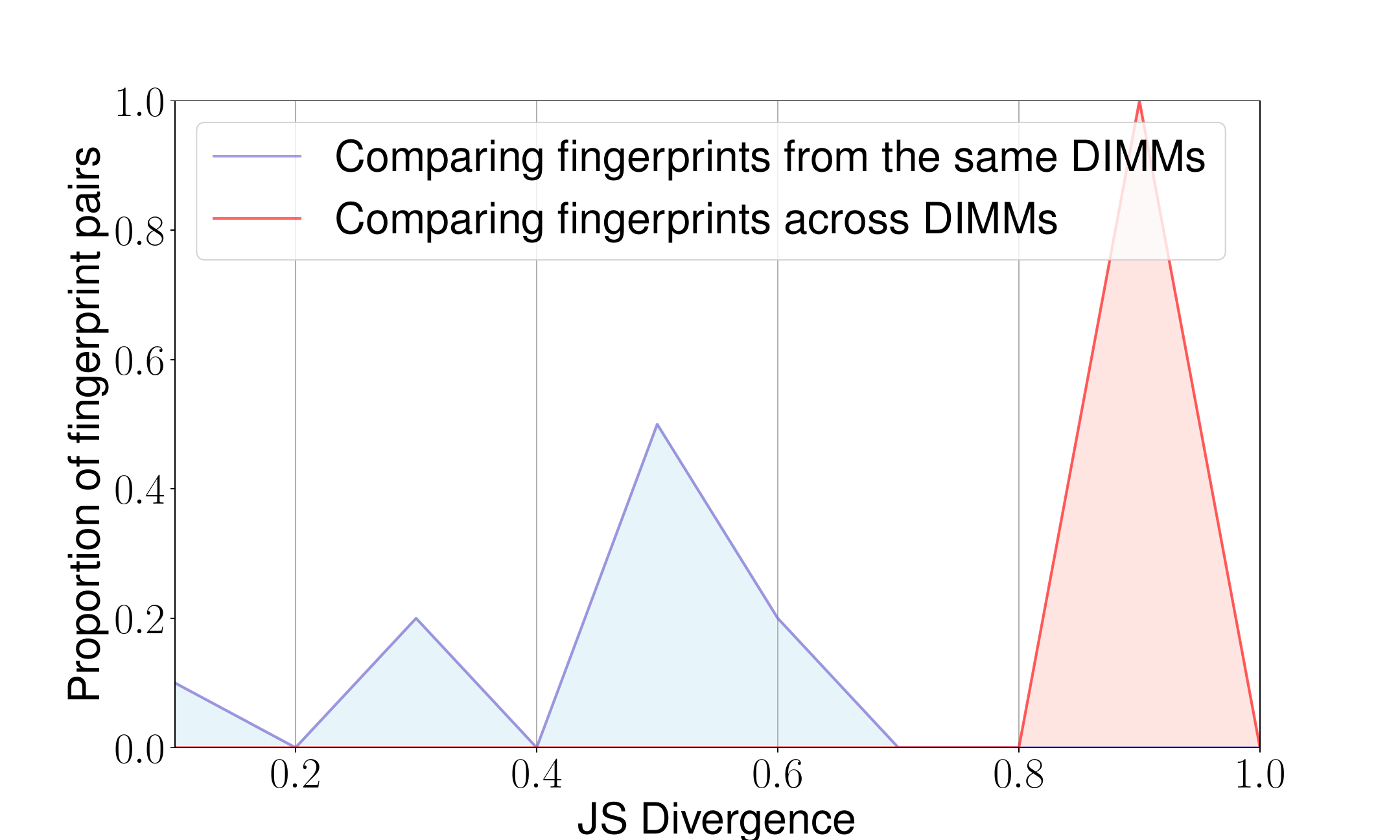}
                }
            \hspace{1.0ex}
            \subfigure[Distribution of JS divergence values across all pairs of 
            fingerprints from 2 identical 1Rx16 DIMMs]
                {
                \label{fig:b_short_1rx16}
                \includegraphics[width=0.3\textwidth]{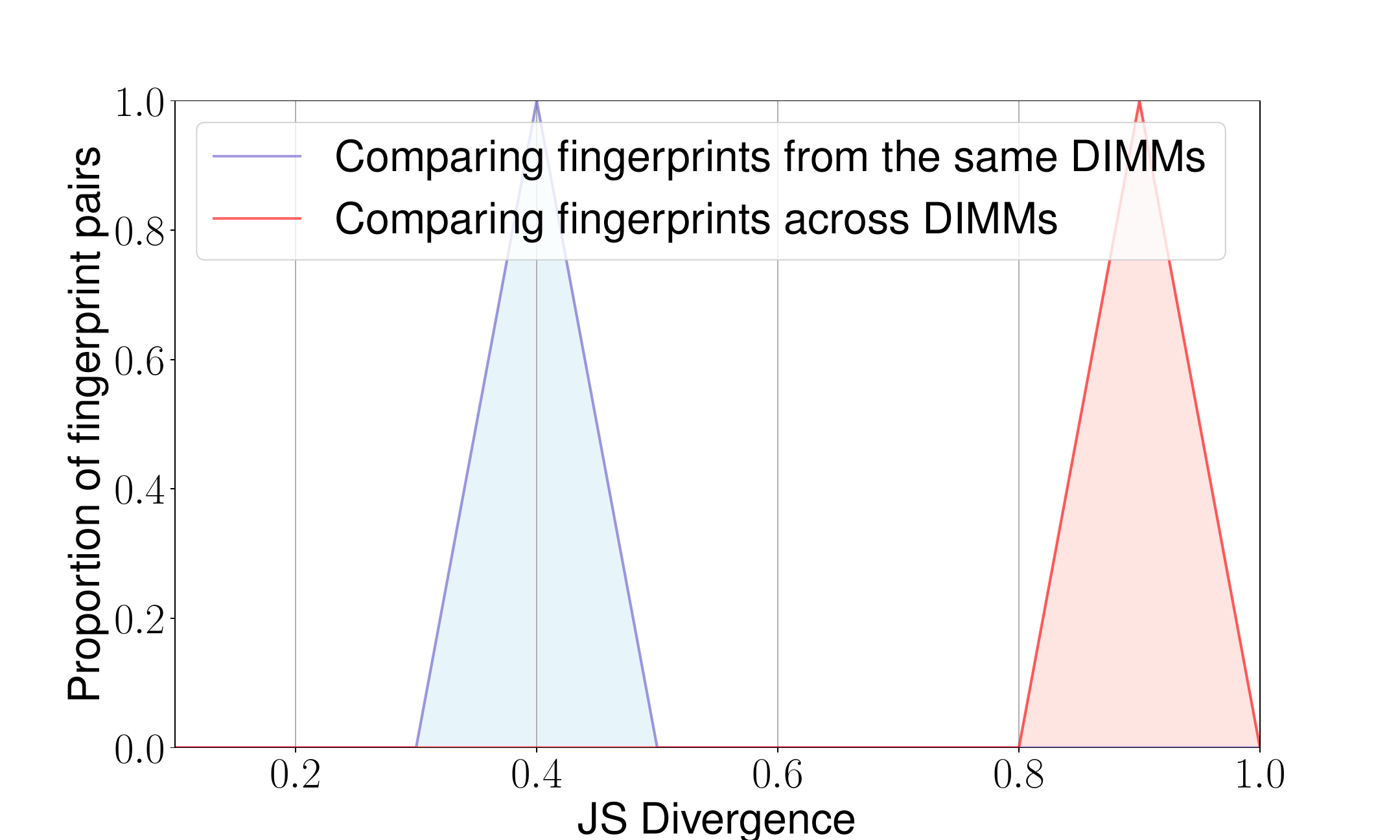}
                }
            }
            \caption{Plots showing the distribution of JS divergence values when
            comparing bit flip distributions obtained from the same pair of DIMMs and
            across different DIMMs.\\}
        \end{center}
    \end{figure*}

    Figures \ref{fig:b_short_2rx8}, \ref{fig:b_short_1rx8} and
    \ref{fig:b_short_1rx16} show the distribution of JS divergence values across pairs of fingerprints from identical 2Rx8, 1Rx8 and 1Rx16 DIMMs, respectively from a 
    different DRAM manufacturer to the one mentioned in \S\ref{sec:eval_uniq}.
    \label{app:alt_manuf}

    \section{Inapplicability of common defenses}
\label{app:inapplicability}
\subsection{Standard fingerprinting defenses}
Standard mitigations against fingerprinting such as normalization 
\cite{fp-stalker} or enforcing permissioned access \cite{battery-fingerprint} 
cannot be employed against \system. Our results demonstrate unique and stable 
fingerprints even among homogeneous devices. Normalization as a defense 
against \system would require eliminating process variation in the 
manufacture of DRAM chips, which is difficult to implement. Since all 
applications including benign applications access memory, blocking access 
to memory or requiring user permission to access memory would not be a 
viable mitigation against \system. To the best of our knowledge, triggering 
additional bit flips to obfuscate or spoof the distributions extracted 
by \system  are not viable since they risk hurting benign memory usage.

\subsection{Standard Rowhammer mitigations}
With \system, we extracted fingerprints on DIMMs that use 
in-DRAM TRR to mitigate Rowhammer. Different hammering patterns 
being able to evade TRR and trigger bit flips on different DIMMs 
shows that fingerprinters can overcome most TRR implementations.
Since we assume that fingerprinters can run experiments on their own
devices to discover ways to trigger bit flips, we anticipate
that they can also overcome other defenses
that attempt to mitigate Rowhammer. For example, researchers have 
explored the possibility of using ECC to defend against Rowhammer.
However, existing research \cite{rowhammer-ecc} has already shown 
that bit flips can also be triggered on ECC equipped systems. Thus, ECC 
does not guarantee a defense against \system. Most recently, researchers
have also managed to overcome in-DRAM ECC \cite{ddr5} to trigger bit flips 
on DDR5 DIMMs \cite{zenhammer}.
    
    \section{DRAM mapping and timing side-channel to determine DIMM geometry}
        \label{app:timing}
\begin{figure}[h]
        \centering
          \includegraphics[width=\columnwidth]{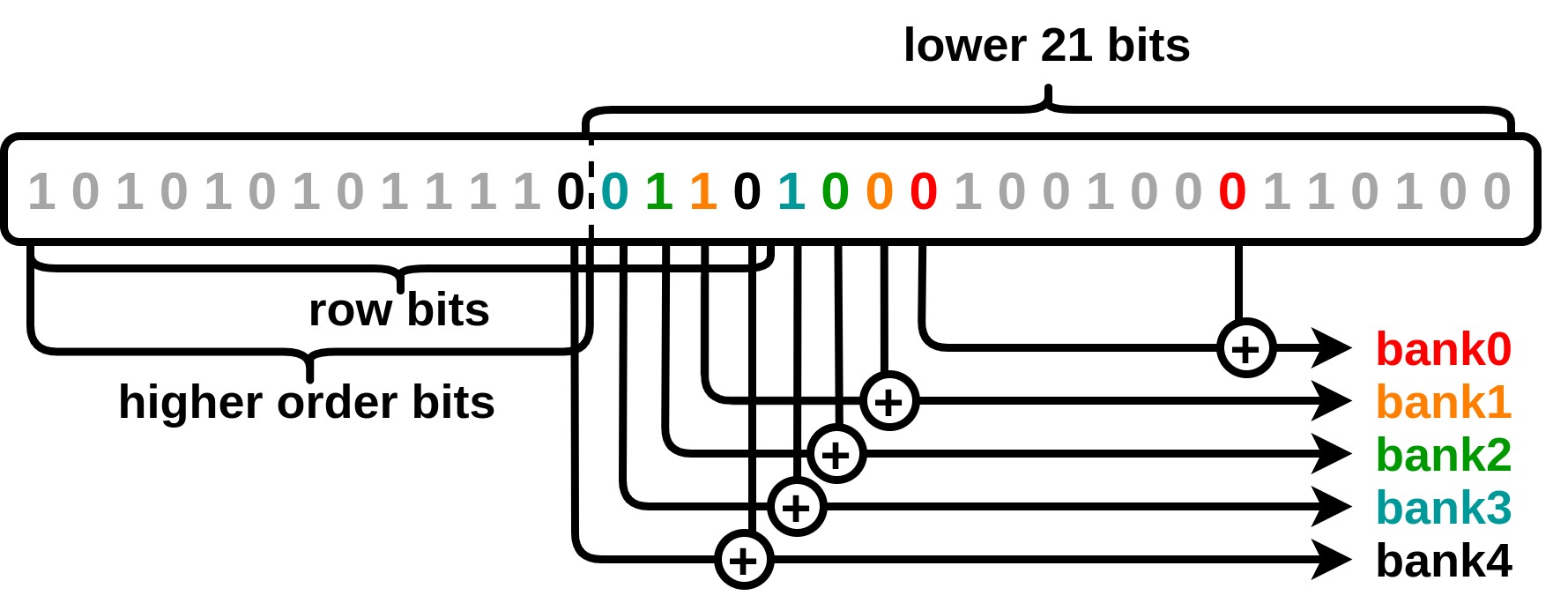}
          \caption{Highlighting the XOR bank functions used on an Intel
          Kaby Lake processor.\\}
          \label{fig:address_2R}
          \hrulefill
        \end{figure}

        We can figure out the geometry of a DIMM from a given address.
        For this example, let us assume that our machine is an Intel
        Core i7-7700 (Kaby Lake) equipped with a 16 GB DIMM, where
        the highest possible address is confined within 34 bits.
        Let’s say that our given physical address is \texttt{0x2abcd1234}.
        Previous research has shown that Intel uses XOR functions to
        decode physical addresses to map it to the device~\cite{drama}.
        We used DRAMA to reverse engineer the mapping functions for our
        machine~\cite{drama,trrespass}. Figure~\ref{fig:address_2R} shows
        the mapping functions look like after decoding the address.
        
        Now, the same information can be obtained from timing side-channels.
        This is partly how DRAMA works as well. This is particularly
        important for us as we need to figure out the number of ranks, banks
        and width of a given DIMM without administrative privileges. Code
        Listing~\ref{code:bank_functions} shows the mapping functions on an
        Intel Kaby Lake with 2 ranks (\texttt{2Rx1}) and 1 rank (\texttt{1Rx1}).
        Keeping our search space limited in this example, let us assume that
        the user's machine is equipped with either of these aforementioned CPUs.
        In order to start the process of finding the number of banks, we first
        need to read the rowbuffer hit time (\textit{t\textsubscript{hit}}). We allocate
        a 2 MB transparent huge page, which gives us access to the lower 21 bits in the physical
        address. These bits can be manipulated from the
        userspace~\cite{smash-rowhammer}. Assume that the starting address is
        \texttt{0x0} and the ending address is \texttt{0xfffff} for the 2 MB chunk.
        
\begin{lstlisting}[language=C++, caption=Bank functions for a Kaby Lake machine with 1 rank and 2 ranks.,label=code:bank_functions]
bank0 = [0x2040, 0x2040];    // 2R or 1R DIMM
bank1 = [0x44000, 0x24000];  // 2R or 1R DIMM
bank2 = [0x88000, 0x48000];  // 2R or 1R DIMM
bank3 = [0x110000, 0x90000]; // 2R or 1R DIMM
bank4 = [0x220000];          // only for 2R DIMM
\end{lstlisting}

        Suppose we access \texttt{0x0} and \texttt{0x1}, we should have
        a rowbuffer hit. Let this time be \textit{t\textsubscript{hit}}.
        Allocated data is interleaved in the
        DRAM~\cite{memorybook,hennesyandpatterson}. In order to cause a
        rowbuffer miss, we need another row from the same bank. This leaves
        us with the fact that we need to find pairs of bits that are XORed
        to determine the bank bits. Column bits take 10 or 11 bits to be
        represented~\cite{dramsim2, dramsim3, ddr4,ddr4-samsung}. The lowest
        3 bits are used to align addresses with byte-sized data. Therefore,
        it is down to the 13th or the 14th bit to represent the first bit for
        a bank function pair (bank0). This is historically seen as true, as it holds
        from 6th generation Intel Core processors to 11th generation Intel 
        Core processors. It's corresponding bit is usually at 6th or 7th bit.
        The important bit pair here is b1, which starts at either 14th or the
        15th bit, depending upon bank0. It's corresponding bit will be at the
        beginning of the row bits. If the DRAM DIMM has 1 rank, then the row
        bits will start at (bank1[0] + 3)th bit. So, we will guess that the row
        bit start at either bit index 17 or 18 and increment by 1. This also 
        implies that we need to set its corresponding XOR bit to 1. This 
        address will be \texttt{0x24000}. If accessing addresses \texttt{0x0}
        and \texttt{0x24000} results in a time say \textit{t\textsubscript{0}},
        such that $t_0 > t_{hit}$ by a \textit{large margin}, then we conclude that
        the DIMM is a single ranked DIMM. If not, then we repeat the process by
        assuming that the starting row bit is at bit index 18 (\texttt{0x0} and
        \texttt{0x44000}). This will imply that we have a DIMM with 2 ranks.
        
        For the case of \texttt{x16}, the starting row index is either 16 or 17
        depending upon the number of ranks. We repeat the aforementioned experiment
        for this case as well. The only point of contention is when we have a
        \texttt{2Rx16} and a \texttt{1Rx8} DIMMs. The starting row index is same, \textit{i.e.} 17.
        In this case we simply run all the known Rowhammer patterns for both the
        cases in the hopes of finding the correct pattern which produces bit flips.
    \section{Extension to configurations with multiple DIMMs across channels}
\label{sec:mult_chan}
If a user's device has multiple DIMMs across channels, contiguous
addresses are interleaved in DIMMs across channels in addition to being 
interleaved across banks. Thus, while manipulating the available 21 bits
in the address of a transparent huge page to set primary aggressors (and to
scan for bit flips), we have to make sure that we do not alter the bits that 
correspond to the channel. For example, in case of 1Rx8 DIMMs on Kaby Lake 
machines, we can alter the row within a bank by modifying bits above the 18th bit 
(bit at index 17 with the least significant bit being bit 0) in the address of the 
huge page. From our experiments with DRAMA~\cite{drama}, we suspect that the channel bits can be derived from a combination of the bits at indices 7, 9, 14, and 17. We suspect that the bits at indices 27 and 28 also contribute to the channel, but they are not relevant since 
they fall outside the bits we can manipulate. 
Thus, we do not change the bit at index 17 to ensure that
we do not change the channel.
    \end{appendix}

\end{document}